\documentstyle[emulateapj,epsf]{article}

\def\gtsima{$\, \buildrel > \over \sim \,$}
\def\ltsima{$\, \buildrel < \over \sim \,$}
\def\simgt{\lower.5ex\hbox{\gtsima}}
\def\simlt{\lower.5ex\hbox{\ltsima}}
\def\sm{$\sim\,$}

\def\onesigma{$1\,\sigma$}
\def\nhat{\ifmmode {\hat{\bf n}}\else${\hat {\bf n}}$\fi}
\def\vtf{v_{{\rm TF}}}
\def\degs{\ifmmode^\circ\else$^\circ$\fi}
\def\kms{\ifmmode{\rm km}\,{\rm s}^{-1}\else km$\,$s$^{-1}$\fi}
\def\kmsmpc{\kms\ {{\rm Mpc}}^{-1}}
\def\etal{{\sl et al.}}
\def\apriori{{\rm a priori}}

\def\h1{\ifmmode h^{-1}\else$h^{-1}$\fi}
\def\dnsigma{$D_n$-$\sigma$}

\def\erf{{\rm erf}}

\def\sigv{\sigma_v}
\def\sigI{\sigma_I}
\def\sigP{\sigma_P}
\def\sigS{\sigma_S}
\def\sigeff{\sigma_{{\rm eff}}}
\def\bfv{{\bf v}}
\def\vev#1{{\left\langle#1\right\rangle}}
\def\call{{\cal L}}
\lefthead{WILLICK}
\righthead{LP10K SURVEY}
\begin{document}
\submitted{Submitted to the Astrophysical Journal}
\title{The LCO/Palomar 10,000 \kms\ Cluster Survey. II.\\ Constraints
on Large-Scale Streaming}

\author{Jeffrey A.\ Willick\altaffilmark{1}} 
\affil{Department of Physics, Stanford University, Stanford, CA 94305-4060 \\
 E-mail: jeffw@perseus.stanford.edu}
\altaffiltext{1}{Cottrell Scholar of
Research Corporation}

\begin{abstract}
The LCO/Palomar 10,000 \kms\ (LP10K) Tully-Fisher (TF)
data set is used to test for bulk streaming motions on a \sm
150\h1\ Mpc scale. 
The sample consists of 172 cluster
galaxies in the original target range of the survey, 
9000--13,000 \kms\ redshift,
plus an additional 72 galaxies with $cz\le 30,\!000\ \kms.$
A maximum-likelihood analysis that is insensitive to
Malmquist and selection bias effects is used to constrain the
bulk velocity parameters, and realistic Monte-Carlo simulations
are carried out to correct residual
biases and determine statistical errors. When the
analysis is restricted to the original target range, the bias-corrected bulk flow
is $v_B=720 \pm 280\ \kms$ (\onesigma\ error)
in the direction $l=266\degs,$
$b=19\degs,$ with an overall \onesigma\ directional error of 38\degs. 
When all objects
out to $z=0.1$ are included the result is virtually unchanged,
$v_B=700 \pm 250\ \kms$ toward $l=272\degs,$ $b=10\degs$
with a directional uncertainty of 35\degs. 
The hypothesis
that the Hubble flow has converged to the
CMB frame at distances $\simlt 100\h1$ Mpc is
ruled out at the 97\% confidence level.
The data are inconsistent with the flow vector
found by Lauer \& Postman; though similar
in amplitude, the LP10K and Lauer-Postman flow
vectors are nearly orthogonal. 
However, the LP10K bulk flow is consistent with that 
obtained from the SMAC survey of elliptical galaxies recently
described by Hudson \etal\ If correct, the
LP10K results indicate that the convergence
depth for the Hubble flow is $\simgt 150\h1$ Mpc.
However, the modest statistical significance of these results,
together with contrasting claims recently made in the literature,
suggest that further observational data are required before firm
conclusions are drawn.
\end{abstract}

\section{Introduction}
For the past decade, establishing
the scale of the largest bulk flows has been one of
the major goals
of observational cosmology. In the late 1980s,
the ``7-Samurai''  (7S) group 
used the \dnsigma\
relation\footnote{\dnsigma\ is one example
of a class of elliptical galaxy scaling relations  
known as the Fundamental Plane (FP). We adopt the latter term
for the remainder of the paper.}
to show that the peculiar velocity
field in the nearby
universe, $cz \simlt 4000\ \kms,$
is dominated by a coherent, large-amplitude ($v_B \approx 500\ \kms$) 
bulk motion in the
direction of the ``Great Attractor'' (GA) near $l=310\degs,$
$b=10\degs$ (Dressler \etal\ 1987;
Lynden-Bell \etal\ 1988). 
Willick (1990) obtained Tully-Fisher
(TF) data for a sample of over 300 field spirals, and found
that the Perseus-Pisces
supercluster, at a distance of \sm 50\h1\ Mpc and
on the opposite side of the sky at $l\sim 110\degs,$
$b\sim -30\degs,$ moves in the same
direction as the 7S ellipticals
at a velocity of \sm 400 \kms. 
A similar result was
obtained by Han \& Mould (1992) from a
cluster TF sample in Perseus-Pisces.
Mathewson \etal\ (1992) analyzed over 1300 Southern-sky
TF spirals, and found that the flow identified 
by the 7S continued beyond the GA, because spirals in
the GA region itself were moving rapidly away from
the Local Group (LG). Courteau \etal\ (1993) used a preliminary
version of the Mark III Catalog of Galaxy Peculiar Velocities
(Willick \etal\ 1997a) to measure the bulk flow for the entire
volume within 60\h1\ Mpc, finding $v_B = 360 \pm 40\ \kms$ toward
$l=294\degs,$  $b=0\degs.$ A recent reanalysis of the
Mark III Catalog using the POTENT method by Dekel \etal\ (1998) finds
$v_B = 370 \pm 110\ \kms$ toward $l=305\degs,$ $b=14\degs$
for the volume within 50\h1\ Mpc. 

Thus, TF and FP data sets acquired through the
early 1990s agreed on the reality of coherent
bulk flows within \sm 50\h1\ Mpc. These motions
were measured relative to the reference frame in which
the dipole anisotropy of the Cosmic Microwave Background
(CMB) vanishes, henceforward the ``CMB'' frame. The
LG itself moves with a velocity of 627 \kms\ toward
$l=276\degs,$ $b=30\degs$ in the CMB frame (Kogut \etal\ 1993).
This direction is within 30--40\degs\ of the observed
bulk flows, suggesting that the LG motion itself
is generated, at least in part, on $\simgt 50\h1$ Mpc scales.

The studies cited above were not, however, deep enough to
establish whether the bulk flows ended, or converged, beyond 50\h1\ Mpc.
Evidence of nonconvergence beyond this distance was first provided by
the work of Lauer \& Postman (1994,
LP94), who used brightest cluster
galaxies (BCGs) as a distance indicator. LP94 analyzed a sample
of 119 BCGs out to a distance of 15,000 \kms, and concluded that
the entire volume out to that distance was moving coherently
at \sm 700 \kms\ in the direction $l=343\degs,$ $b=52\degs.$ This flow
vector was about $60\degs$ away from
the motions detected in the earlier FP and TF-based studies,
and was on a much larger scale. 

However, a number of recent studies have challenged 
the validity of both the LP94
result in particular, and very large-scale
($\simgt 100\h1$ Mpc) bulk flows in general. 
Giovanelli \etal\ (1996) analyzed a large sample
of TF spirals toward the apex and antapex of the LP94 flow
direction, and found zero net peculiar velocity along that
axis out to a distance of \sm 70\h1\ Mpc. Riess, Press, \& Kirshner
(1995) used 13 Supernovae of Type Ia (SN Ia), each with a distance
error less than 10\%, to constrain the magnitude and direction
of the bulk flow within 10,000 \kms\ redshift. They showed that their
data set was inconsistent with the LP94 flow vector at
99\% confidence.
More recently, Giovanelli \etal\ (1998a,b) and
Dale \etal\ (1998) have analyzed I-band field and cluster
TF samples to estimate the convergence scale. Giovanelli
(1998a,b) found convergence at $\simlt 6000\ \kms$ using
primarily field spirals. Dale \etal\ (1998) combined
the distant ($cz>4500\ \kms$) portion of the Giovanelli (1998a,b)
sample with 522 spirals in 52 Abell clusters at distances
between \sm 50 and \sm 200\h1\ Mpc. The effective depth
of this combined sample was \sm 9500 \kms.
Dale \etal\ found a bulk
velocity consistent with zero, and at most 200 \kms,
for this volume. 
The EFAR group (Wegner \etal\ 1996,
1998) has obtained FP data for \sm 500 ellipticals in 84 clusters
in two patches of the sky. They also find generally small cluster
peculiar velocities in the mean, and in particular rule
out the LP flow at 99\% confidence (Saglia \etal\ 1998).
However, the limited sky coverage of the EFAR sample means
that it is not sensitive to the full range of possible flow
directions.

In contrast, the recently completed SMAC survey of
elliptical galaxies
(Hudson \etal\ 1998a,b) has found evidence for a large-scale
bulk flow, though not in the LP94 direction.
The SMAC group measured FP distances for 697 early-type
galaxies in 56 clusters with $cz\leq 12,000\ \kms,$
and found a bulk
flow of $640 \pm 200\ \kms$ in the direction
$l=260 \pm 15\degs,$ $b=-1 \pm 12\degs.$ This flow vector is 
within 40--50\degs\ of, and is similar in amplitude to, the motions detected
5--10 years ago by the 7S, Willick (1990), 
Mathewson \etal\ (1992), and Courteau \etal\ (1993). 
However, the SMAC data set
has about twice the effective depth of those earlier surveys,
and thus suggests that the convergence
scale could be $\simgt 8000\ \kms.$ 

In short, while a number of studies agree on the
reality of a significant bulk flow within 50\h1\ Mpc,
the persistence of
such motions to distances beyond \sm 80\h1\ Mpc  remains
controversial. The purpose of this paper is to address this
issue using a new data set. The outline of the paper is as
follows. In \S~2, we describe the new TF data set, known
as the LP10K sample. In \S~3, we describe the maximum-likelihood
method used to constrain the bulk flow vector. In \S~4,
we apply this method to the LP10K data set.
In \S~5, we describe
Monte-Carlo simulations of the sample, and discuss how these simulations
are used to assess the statistical significance of the results.
Finally, in \S~6 we further discuss
and summarize the main results of the paper.

\section{The LP10K Survey}
In early 1992 the author initiated a survey designed specifically
to test whether the bulk streaming observed within
50--60\h1\ Mpc persists to distances
$\simgt 100\h1$ Mpc. The survey targeted spiral and
elliptical galaxies in 15 Abell clusters
with published redshifts in the range $9000 \le cz \le 12,\!000\ \kms,$
and utilized the TF and FP relations as distance indicators.\footnote{As 
discussed below, the upper end of the redshift range for
these clusters turns out to be closer to 13,000 \kms, the value quoted
in the Abstract.}
To maximize sensitivity to a bulk flow,
the 15 clusters were selected to be distributed
as isotropically as possible on the sky. Limitations to isotropy
were imposed only by the requirement of Galactic latitude $|b| \ge 20\degs$
to minimize extinction effects and
by the finite number of clusters observed. 
Southern-sky clusters were observed
from the Las Camapanas Observatory (LCO), while Northern-sky clusters were
observed from Palomar Observatory. The survey is thus known as
the LCO/Palomar 10,000 \kms\ Cluster Survey (LP10K).

The LP10K observing  runs spanned the period 1992 March--1995 September,
and totaled over 100 nights of observations. The observing strategy, 
data reduction methods, and
the modeling of the TF relation were described in detail by
Willick (1999a), hereafter Paper I. 
Although
the LP10K TF data set is fully analyzed, observations and reductions
of the elliptical galaxy portion of the survey are ongoing, with
completion expected in 2--3 years. {\em The results presented here
are derived from TF data only.} The full TF data set, along with
a variety of tests of the accuracy and repeatability of the observations,
will be presented in the third paper in this series (Willick 1999b, Paper III).

Figure~1 shows the sky positions of the 15 LP10K clusters
in Galactic coordinates. (The equatorial coordinates of
the clusters are listed in Table 1 of Paper I.) 
The clusters are seen to be well distributed around the 
sky above $|b|=20.$ There are
clusters near both the apex and the antapex of the
CMB dipole at $l=276\degs,$ $b=30\degs,$ ensuring sensitivity
to flows along this axis. The axis corresponding to
the LP94 motion is similarly well-sampled.  An unexpected
feature of the survey, discussed in Paper I, was that many TF
galaxies turned out to have redshifts well in excess
of the published cluster velocities.
To appreciate the extent of this effect, Figure~1 indicates
the mean redshift $\vev{cz}$ of each cluster TF sample by point type:
solid circles indicate mean redshifts in the original target
range, $\vev{cz} \leq 12,\!000\ \kms$ (in practice all of
these have $\vev{cz} > 11,\!000\ \kms$ as well); open circles
show clusters with $12,\!000 < \vev{cz} \leq 15,\!000$; and
starred symbols clusters with $\vev{cz} > 15,\!000\ \kms.$
It is apparent that a sizable {\em majority\/} of LP10K sample
clusters have mean redshifts greater than 12,000 \kms,
the upper limit of the original target range. Indeed, the
sample includes a significant number of objects
with $z \simeq 0.1.$ The number of TF galaxies per cluster
is denoted by point size. The cluster TF sample sizes range
from 8 to 26 objects, with an average of about 16 galaxies
per cluster.

Table 1 provides additional information on the cluster TF
samples. Column 1 gives the cluster name according
to the catalogue of Abell, Corwin, \& Olowin (1989; ACO). 
Columns 2 and 3 give the cluster Galactic coordinates $l$ and $b.$
There follows a listing of three redshifts for the cluster: column
4 lists an updated
value  derived from a literature search, typically weighted
by early-type galaxies in the cluster core; column 5 lists
the mean redshift
of all LP10K TF galaxies in the cluster field; and column 6 
lists the mean sample redshift when
only objects with $7000 \leq cz \leq 15,000\ \kms$ are included
(the significance of this redshift cut is discussed below).
All redshifts are given with respect to the CMB frame.
Column 7 lists the total number of TF galaxies in each
cluster field; these values correspond to the point sizes
in Figure~1 as well as the redshifts in column 5. Column 8 lists the number
of galaxies with $7000 \leq cz \leq 15,000\ \kms$ and corresond
to the redshift in column 6.
Column 9 provides keys
to the sources from which the literature-based redshift
are derived.
\vbox{%
\begin{center}
\leavevmode
\hbox{%
\epsfxsize=8.9cm
\epsffile{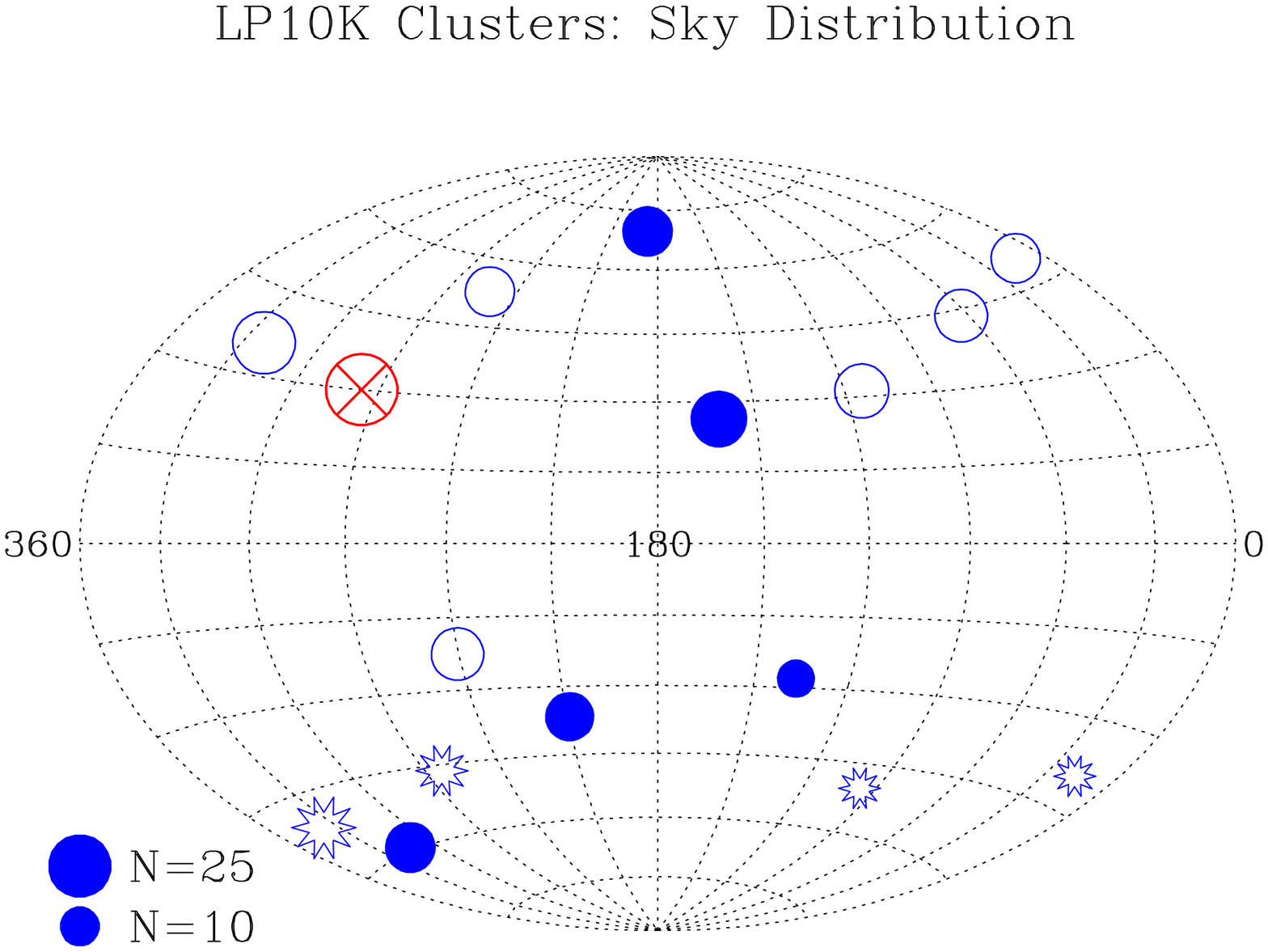}}
\begin{small}
\figcaption{%
Sky positions of the fifteen LP10K clusters. Filled circles
represent clusters for which the mean redshift of
the LP10K TF sample is $\leq 12,\!000\ \kms.$ 
Open circles represent
clusters with mean TF sample redshift $12,\!000\ \kms \leq
cz \leq 15,\!000\ \kms,$ and starred symbols those
with mean $cz > 15,\!000\ \kms.$ The number
of galaxies in the cluster TF sample is symbolized by
the point size, as indicated by the key at the lower left.
The ``{\large $\times$}'' enclosed by a circle shows the direction of
the LG motion with respect to the CMB.
\label{fig:LKP10Kdist}}
\end{small}
\end{center}}

The prevalence of relatively high-redshift objects
in the LP10K TF data set is a consequence of
the way sample galaxies were selected, namely, on the
basis of magnitude and morphology from CCD images
(cf.\ Paper I), and proximity to the cluster center,
not on prior knowledge of the redshift.
Most sample galaxies, in fact, did not have a measured redshift
prior to the LP10K observations. The high-redshift galaxies---which
we define here, somewhat arbitarily, as those with $cz>15,000\ \kms$---present
a problem only insofar as they reduce the number of objects
in the 9000--12,000 \kms\ redshift shell at which
the survey was originally targeted.  On the other hand,
the presence of higher-redshift galaxies 
provides some sensitivity to a bulk flow on scales
considerably larger than the targeted shell. 
\begin{table*}
\begin{minipage}{180mm}
\caption[ ]{\centerline{{\footnotesize
LITERATURE, FULL SAMPLE, AND EXTENDED TARGET RANGE REDSHIFTS$^a$}}}
\centerline{
\begin{tabular}{crrrrrrrl}
\hline\hline\multicolumn{7}{c}{} \\
Cluster&\multicolumn{1}{c}{$l$}&\multicolumn{1}{c}{$\;b$}&$cz_{{\rm LIT}}$&
$cz_{{\rm FS}}$&$cz_{{\rm ETR}}$&\multicolumn{1}{c}{$N_{{\rm FS}}$}&
\multicolumn{1}{c}{$N_{{\rm ETR}}$} & References$^b$ \\ \hline 
A0260&$137.1$&$-28.1$&10621&12046&10505& 8& 7& 4,7 \\
A0496&$209.3$&$-36.6$& 9831&11215&10124&15&11& 4,6,7\\
A0576&$161.3$&$ 26.3$&11493&11973&11973&20&20& 4,6\\ 
A1139&$251.7$&$ 52.7$&12265&13491&12310&15&14& 2,4\\
A1228&$186.7$&$ 69.6$&11217&11261&11413&16&13& 2,4\\
A2063&$ 12.7$&$ 49.8$&10803&12546&10776&14&12& 4,6,7\\
A2199&$ 62.9$&$ 43.8$& 8989&12961& 9353&17&12& 4,6,7\\
A2247&$114.4$&$ 31.2$&11472&12881&11360&18&13& 4,6,7\\
A2657&$ 96.7$&$-50.3$&11636&17664&11787&10& 3& 4,6,7\\
A2731&$313.8$&$-59.3$& 9246&11090& 9161&16&14& 1,5\\
A3202&$263.5$&$-46.3$&11600&16409&12481&17& 8& 1\\
A3381&$240.4$&$-22.5$&11517&13028&10256&17&12& 1,3,5\\
A3578&$321.7$&$ 35.8$&12269&13074&12840&24&18& 1,4\\
A3733&$ 17.5$&$-39.5$&10791&16257&10903&11& 5& 5\\
A3869&$336.5$&$-51.2$&11839&19485&12450&26&10& 5\\ \hline\hline
\end{tabular}
}
\smallskip

Notes: (a) All redshifts are listed as CMB frame radial
velocities in \kms. 
(b) The entries in the References column are keys
to the source(s) of the literature redshift, as
follows: (1) Abell, Corwin, \& Olowin (1989, ACO). 
(2) Dale \etal\ 1997; 
(3) Jorgensen \etal\ 1996. (4) Postman, Huchra,
\& Geller 1992. (5) Postman \& Lauer 1995. (6) Struble \& Rood 1991.
(7) Wegner \etal\ 1996.
\end{minipage}
\end{table*}

To clarify the analysis presented in \S~3, it will prove
useful to distinguish between the full LP10K TF
sample and a subsample that maximizes sensitivity
to the originally targeted redshift shell.
We construct this subsample by excluding all
galaxies with $cz<7000\ \kms$ and $cz>15,000\ \kms.$ This
cut yields the cluster
redshifts and sample sizes given in columns 6 and 8 
of Table~1. We henceforward refer to this subsample as the
``extended target range'' (ETR).  The ETR subsample 
comprises 172  galaxies, most of which are bona fide
cluster members, judging from the fairly good agreement
between $cz_{{\rm LIT}}$ and $cz_{{\rm ETR}}$ given in Table~1.
The notable exceptions to this trend are the clusters A3202
and A3381, for which the ETR redshifts are, respectively,
\sm 900 \kms\ greater and \sm 1300 \kms\ less than the literature values.
A3202 is, however, the cluster for which the literature value
is the most uncertain, and we are inclined to regard the
the LP10K ETR redshift as more accurate. 
For A3381 the source of the discrepancy is probably
significant subclustering within the cluster; the LP10K
TF sample simply weights a lower-redshift clump more strongly
than the studies from the literature.

In summary, it is reasonable
to conclude that the ETR subsample is representative
of the fifteen LP10K clusters {\em per se.} 
This is not to say
that all ETR galaxies are members of the virialized cluster cores;
indeed, as shown in \S~3.2.2, most almost certainly are not, because
a Hubble expansion plus bulk flow model turns out to be a much better
description of the TF data than a model which assumes all members
of a given cluster are equidistant. A second point is that
the ETR shell appears to be better described as sampling
the redshift range 9000--13,000 \kms, as compared with
the original upper limit of 12,000. 

Distinct from the ETR is the entire LP10K TF data set 
irrespective of redshift, henceforward the
``full sample'' (FS) . 
The FS comprises 244 galaxies.
Sixty-four of the 72 galaxies not in the ETR 
have redshifts $>15,000\ \kms,$ while the remaining eight
have redshifts between 5000 and 7000 \kms. 

\section{Method}
As in Paper I, we use a maximum-likelihood method based on
the inverse TF relation, but with two key differences. First,
we predict distances from redshift and position on the sky
using a
Hubble expansion plus bulk flow model. (In Paper I 
bulk flow was omitted.)
The distance in Mpc to the
$i$th galaxy is given by
\begin{equation}
d_i = H_0^{-1} \left[cz_i - \bfv_p \cdot \nhat_i \right]\,,
\label{eq:di}
\end{equation}
where $\nhat_i$ is a unit vector
in the direction of the galaxy, and $cz_i$ its CMB frame redshift in \kms.
(We also consider, and then reject, a model in which $cz_i$
is replaced by the mean redshift of the cluster to which
the galaxy nominally belongs; cf.\ \S~4.2 for details.)
One or more components of the bulk flow vector $\bfv_p$ may
be treated as free parameters in the maximum likelihood fit,
but $\bfv_p$ is the same for every
galaxy in the sample.
As in Paper I we take $H_0=65\ \kmsmpc;$ the adopted value
affects only the zero point of the TF relation, not the derived bulk
flow vector.

Given $d_i,$  the predicted value of the
circular velocity parameter is calculated as
\begin{equation}
\eta_{i, pred} = -e(m_i - 5\log d_i -25 -D) + \alpha\mu_i +\beta c_i
\label{eq:etapred}
\end{equation}
where $m_i,$ $\mu_{e,i},$ and $c_i$ are the observed apparent magnitude,
effective central surface brightness, and concentration index 
of galaxy $i,$ respectively (cf.\ Paper I for details). 
The quantities $D$ and $e$
are the zero point and slope of the TF relation, while $\alpha$
and $\beta$ are additional TF parameters whose
significance is discussed in Paper I.
The other important quantity that enters the likelihood analysis is
the observed value of the circular velocity parameter, $\eta_i.$
Recall from Paper I that $\eta_i$ is defined in terms of an additional
TF parameter, $f_s,$ which determines the radius at which the
rotation velocity is evaluated from the full RC. This is known
as the ``$f_s$-formulation'' of the TF relation (Paper I), and
is the approach used here. (An alternative approach
discussed in Paper I, the  ``$x_t$-formulation,'' is
essentially equivalent and is not used in this paper.)

The second key difference from  Paper I
is that we now properly take into account the statistical
effects of having multiple photometric
and/or spectroscopic measurements for a large number
of individual galaxies. In Paper I we included each
photometry/spectroscopy data pair for a particular
galaxy as a ``data object,'' and all such objects entered
equally into the likelihood computation. In doing so we neglected
the correlations among the various data objects derived
from a single galaxy, but crudely estimated their effect
by scaling the likelihood statistic by the ratio
of distinct galaxies to the total number of data objects. This
was a conservative approach, as the actual number of degrees
of freedom is greater than the number of distinct galaxies because
the repeat measurements are partially independent. Such an approach
was acceptable in Paper I because the effects we sought
to demonstrate had very strong statistical significance.

The bulk flow parameters, however, have relatively
large errors, and it is thus important to employ
a rigorous rigorous likelihood statistic. This is done as follows.
Suppose that the $i$th distinct galaxy was observed only once
spectroscopically, yielding circular velocity parameter
$\eta_i,$ but has $n_i$ photometric measurements
(in practice, $n_i\le 4$), yielding apparent magnitudes
$m_{ij},$ $j=1,...,n_i.$ Then the conditional probability density
for the observed velocity width, given the $n_i$ photometry
measurements and the redshift is
\begin{equation}
P(\eta_i|m_{ij},cz_i) =\frac{1}{\sqrt{2\pi\sigeff^2}}
\exp\left\{-\frac{\left[
\eta_i - \eta_{i,pred}({\overline m_i})\right]^2}{2\sigeff^2}\right\}\,,
\label{eq:peta1gm}
\end{equation}
where:
\begin{enumerate}
\item $\eta_{i,pred}({\overline m_i})$ is the predicted circular velocity parameter
based on the average of the photometric measurements,
${\overline m_i} = \sum_j m_{ij}.$ Because we use a multiparameter
TF relation, equation~\ref{eq:etapred}, ${\overline m_i}$ symbolizes 
averages over all relevant photometric quantities, not only
apparent magnitude.
\item The observed circular velocity parameter $\eta_i$ is also derived
by combining the spectroscopic measurement with average values
of the needed photometric parameters: the inclination $i$ 
and the effective exponential
scale length $r_e$ (cf.\ Paper I). 
\item The effective scatter is computed as the quadrature sum of
its four contributions:
\begin{equation}
\sigeff^2 = \left(\frac{5e}{\ln 10}\,\frac{\sigv}{d_i}\right)^2 + \sigI^2
+ \sigS^2 + \sigP^2/n_i
\end{equation}
where:
\begin{enumerate}
\item $\sigv=250\ \kms$ is, as in Paper I, the assumed velocity noise
relative to the bulk flow model;
\item $\sigI$ is the intrinsic scatter in the inverse TF relation,
taken as a free parameter in the model;
\item $\sigS$ is the portion of the $\eta$-error due to spectroscopic
measurement errors only. It was modeled as
$\sigS = \delta v_{rot}/(\vtf\sin i),$ where $\delta v_{rot},$ treated as a
free parameter in the model, is the circular velocity measurement
error, assumed constant, and $\vtf$ is the predicted value of
the circular velocity. The motivation for this error
model was discussed in Paper I. (The projection factor $\sin i$ was erroneously
neglected in the Paper I analysis.)
\item $\sigP$ is the rms error in measuring and predicting $\eta$
due to photometric measurement errors. Thus, it includes the effects of inclination
and effective radius measurement errors on the measured $\eta,$
as well as those of magnitude, surface brightness, and concentration
index errors on the predicted $\eta.$ The
various photometric errors were assessed
by comparing values for multiply observed objects, and will be
described fully in Paper III. In general, $\sigP$ constitutes a very
small fraction of the overall error budget.
\end{enumerate}
\end{enumerate}
Note that equation~\ref{eq:peta1gm} reduces to equation 10 of
Paper I when $n_i=1.$

If galaxy $i$ has two spectroscopic measurements yielding
circular velocity parameters $\eta_{i1}$ and $\eta_{i2},$
then another term is present in the likelihood, as follows:
\[ P(\eta_{i1},\eta_{i2}|m_{ij},cz_i) = \frac{1}{\sqrt{4\pi\sigS^2}}\,\exp\left\{-\frac{(\eta_{i1}-\eta_{i2})^2}
{2(\sqrt{2}\sigS)^2}\right\}  \]
\begin{equation}
\times \frac{1}{\sqrt{2\pi\sigeff^2}}
\exp\left\{-\frac{\left[
\overline\eta_i - \eta_{i,pred}(\overline m_i)\right]^2}
{2\sigeff^2}\right\}\,,
\label{eq:peta2gm}
\end{equation}
where ${\overline \eta_i}=(\eta_{i1}+\eta_{i2})/2,$ and the
effective scatter is now given by
\begin{equation}
\sigeff^2 = \left(\frac{5e}{\ln 10}\,\frac{\sigv}{d_i}\right)^2 + \sigI^2
+ \sigS^2/2 + \sigP^2/n_i\,.
\end{equation}
Note that the two-spectroscopic measurement likelihood differs
from the single measurement case by the presence of
a term measuring the probability that the two measurements
will differ by the observed amount. The corresponding term
for photometric measurement differences does not
occur because we are using conditional probabilities of
$\eta$ given $m.$ 
The presence of this term gives the
likelihood analysis greater leverage on the velocity
measurement error term $\delta v_{rot}$ than the Paper I
approach, which relied solely on TF residuals to
estimate $\delta v_{rot}.$ In fact, as will be seen,
this leads to a \sm 10\% increase in the estimated $\delta v_{rot},$
a relatively small change. The better-determined $\delta v_{rot}$
in turn means that the resultant value of $\sigI,$
the intrinsic TF scatter, is also more reliably
determined. Our inclusion of photometric measurement
errors, neglected in Paper I, also improves the
$\sigI$ estimate. Thus, our present approach
affords one of the most accurate estimates 
to date of the
the intrinsic TF scatter, a quantity
of considerable interest for galaxy formation theory 
(e.g., Steinmetz \& Navarro 1998). 

The overall likelihood for the sample is given by
${\cal P} = \prod_i P(\eta_i|m_i,cz_i),$ where
the product runs over all distinct galaxies
in the sample (as opposed to all data objects as in Paper I),
and $P(\eta_i|m_i,cz_i)$ is calculated from equation~\ref{eq:peta1gm}
or equation~\ref{eq:peta2gm}, as appropriate. 
In practice, likelihood maximization is achieved
by minimizing the statistic $\call = -2\ln{\cal P}.$
In all cases, $\call$ is minimized with
respect to the various TF parameters (including
those characterizing the TF errors), as well
as the bulk flow components.

Although the maximum likelihood method is effective for
simultaneously determining bulk flow and TF parameters, it
does not yield a rigorous measure of goodness of fit (cf.\
the discussion by Willick \etal\ 1997b). For this
purpose, it is useful to define ``cluster-$\chi^2$'' statistic 
sensitive only to the bulk flow parameters,
and changes in which from one model to the next can
be used to gauge improvements in the fit.
We do so by computing an average radial peculiar velocity
$v_p$ and
corresponding error $\delta v_p$ for each cluster, and comparing
with the predicted radial peculiar velocity $u=\bfv_p\cdot\nhat$ from
the flow model:
\begin{equation}
\chi^2_{{\rm clust}} = \sum_{i=1}^{15} \left( \frac{v_{p,i} - u_i}{\delta v_{p,i}}
\right)^2 \,,
\label{eq:chi2clust}
\end{equation}
where the sum runs over the fifteen LP10K clusters. In calculating
the $\delta v_p,$ we take the scatter contributions $\sigI$ and $\sigS$
to have fixed values, rather than the best-fit values from the maximum
likelihood solution. This ensures that $\chi^2_{{\rm clust}}$ 
measures only bulk flow errors, not differences
in the TF scatter from model to model. The adopted values
are  $\sigI=0.025$
and $\delta v_{rot}=17.5\ \kms,$ close to the
values obtained from the best ETR fit (cf.\ \S~4.4). 

\section{Results}
Before proceeding to the results of the flow analysis,
we address two key technical issues: 
Galactic extinction and the redshift-distance
model.

\subsection{Burstein-Heiles versus Schlegel-Finkbeiner-Davis extinctions}

Two all-sky Galactic extinction maps are presently available: the older Burstein-Heiles
(Burstein \& Heiles 1978, 1982; BH) 
maps, which are based on 21 cm column density and faint galaxy counts, 
and the recently completed Schlegel, Finkbeiner, \& Davis (1998; SFD) maps,
based on DIRBE/IRAS measurements of diffuse IR emission.
The SFD extinctions have been favored in several recent analyses,
and indeed were used in Paper I. Unlike BH, the SFD extinctions
are based directly on dust emission and have comparatively
high spatial resolution. However,
it has not been established beyond doubt that they are free
of systematic errors, such as could arise from the presence of
cold dust invisible to IRAS. The BH extinctions are also vulnerable
to possible systematic effects, such as a variable dust-to-gas
ratio and galaxy count fluctuations. Thus it seems prudent
to use both methods, or linear combinations of them,
and see what effect this has on the results.

We have run the likelihood and cluster-$\chi^2$ 
analyses, for both the ETR and FS samples, correcting
apparent magnitudes and surface brightnesses using BH, SFD, and their
direct average, (BH+SFD)/2. The results 
are given in Table~2, of which a full description is given below; here
we summarize only the conclusions with regard to extinction.
First, the derived
flow vector is quite insensitive to which extinction method
is used. The flow direction shifts about $+20\degs$
in longitude and $+10\degs$ in latitude, and the flow amplitude increases
by a few percent, when SFD extinctions are replaced by
BH extinctions. These changes are smaller than the $1\sigma$
errors, and thus statistically insignificant. The 
cluster-$\chi^2$ statistic is also relatively insensitive
to the extinction method. For the ETR the SFD extinctions
produce a smaller $\chi^2_{{\rm clust}},$ whereas for the FS the situation
is reversed. For both the ETR and the SF,
average extinctions, (BH+SFD)/2, yield likelihood and 
$\chi^2_{{\rm clust}}$ statistics as good, or nearly so, as the better
of SFD or BH. We conclude that the flow analysis does
not provide clear evidence for the superiority of one extinction
scheme over the other, and indicates that averaging
them may produce most reliable results overall.  
We thus adopt the average extinctions to obtain the
final flow vectors for the ETR and the FS, as well
as to compare with alternative solutions such as flow
along the LP94 direction or no flow (see below).

\subsection{Distance assignments}

The LP10K sample consists, nominally, of cluster galaxies.
It is conventional to assume that all members of a given
cluster lie at a common distance, regardless of redshift
(the ``cluster paradigm'').
In Paper I, however, we modeled all galaxy distances by the
Hubble law, thus assuming that even within a cluster there
is a redshift-distance correlation. Such an approach
is obviously called for when analyzing the full sample, 
with its many galaxies well in the background of the targeted clusters.
On the other hand, the
ETR cluster subsamples have mean
redshifts close to published
values for the cluster cores, suggesting that 
the redshift-distance relation for the ETR might be
more aptly modeled by the cluster paradigm. However, some previous
cluster TF studies have shown that even spirals
near cluster cores, as judged by position in redshift space,
may exhibit a redshift-distance relation close to pure Hubble expansion
(Bernstein \etal\ 1994; Willick \etal\ 1995).
The LP10K clusters are too distant, given the
large TF errors, to judge which model is better for a given
cluster. However, we can address this question by carrying
out a cluster paradigm fit to the ETR subsample.
To do so we model
the distance to the $j$th galaxy in the $i$th
cluster by 
\begin{equation}
d_{ij} = H_0^{-1} \left[ \vev{cz_i} - \bfv_p\cdot\nhat\right]\,,
\label{eq:dij}
\end{equation}
where $\vev{cz_i}$ is the mean redshift of the $i$th cluster.
We take this to be the literature-based redshift, column (4)
of Table 1, except in the cases of A3202 and A3381, where
we adopt the mean LP10K ETR redshift, column (6) of Table 1
(see the discussion in \S~2). This distance model implicitly
assumes that any redshift dispersion within a given cluster
is due to virial velocities, not to Hubble expansion.

We applied the maximum likelihood algorithm to the LP10K
ETR subsample using equation~\ref{eq:dij} to assign distances,
and calculated the resultant $\chi^2_{{\rm clust}}.$ The results
are given in row (4) of Table~2. The likelihood statistic is
larger, by 11 units, than the best-fit value assuming free expansion.
Given that the models have the same numbers of degrees of freedom,
this is a highly significant (\sm $3.3\,\sigma$) difference. Similarly,
$\chi^2{{\rm clust}}$ is 7.3 units larger for the cluster paradigm
fit than for the free expansion fit. These statistics demonstrate
that, overall, the redshift-distance relation for
the ETR subsample is much better modeled
by free expansion than the cluster paradigm.

\vbox{%
\begin{center}
\leavevmode
\hbox{%
\epsfxsize=8.9cm
\epsffile{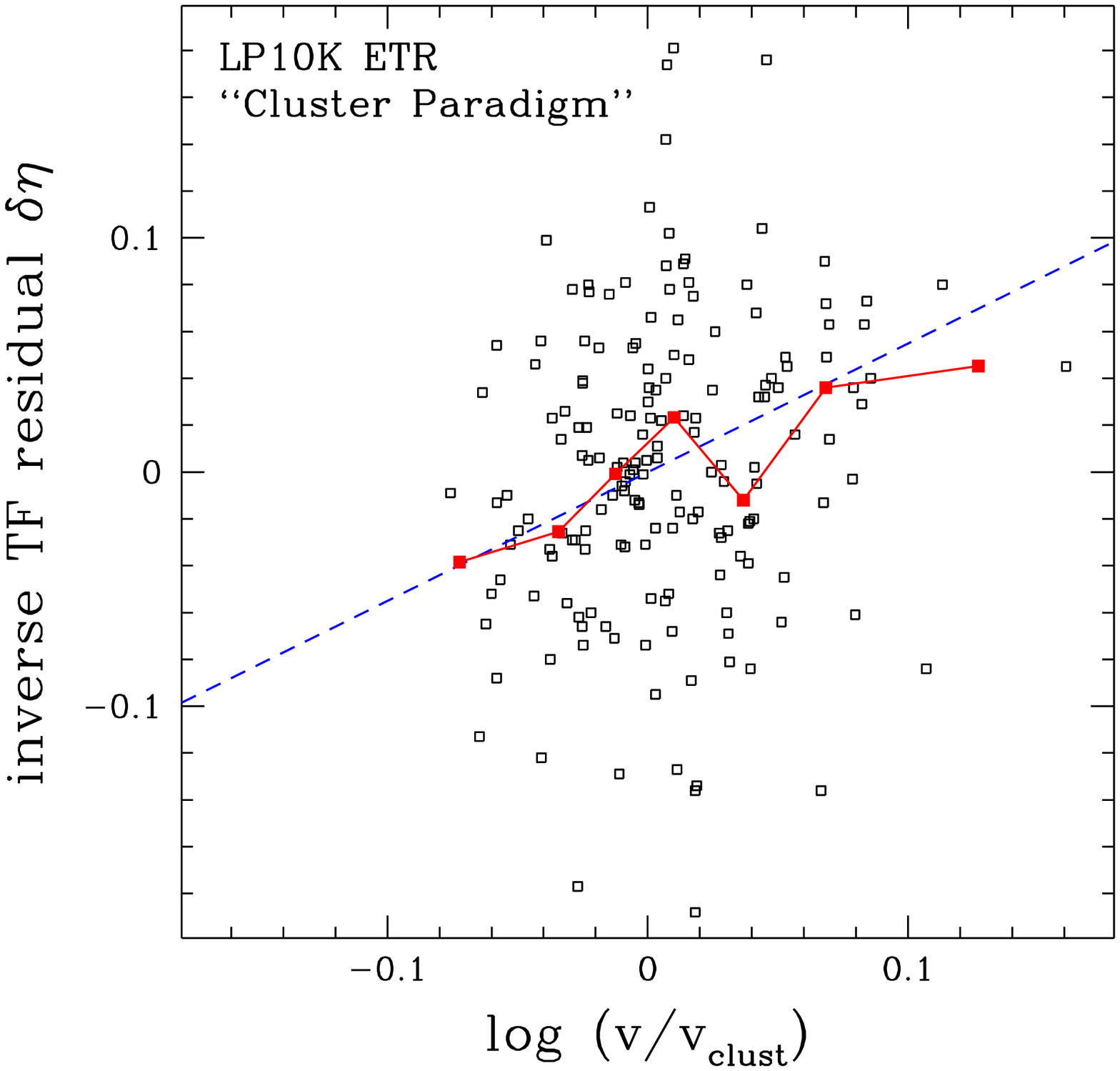}}
\begin{small}
\figcaption{%
Inverse TF residuals from the ETR ``cluster paradigm'' fit,
plotted against $\log(v/v_{clust}),$ where $v$ is the
individual galaxy redshift and $v_{clust}$ is the
mean redshift of its parent cluster. If the galaxies
are better described by free expansion than the
cluster paradigm, the residuals are expected to
follow the relation $\delta\eta \approx 5 e \log(v/v_{clust}),$
shown as the diagonal dashed line in the figure. The
connected filled squares show a running median of
the residuals, which are seen to follow the expansion
relation rather closely. This suggests that the redshift-distance
relation for the LP10K TF spirals is, to within the
accuracy of the data, close to the
free expansion expectation.
\label{fig:etr_cp}}
\end{small}
\end{center}}

This point is demonstrated graphically in Figure~\ref{fig:etr_cp},
where we plot inverse TF residuals from the ETR cluster paradigm
fit versus logarithmic redshift differences between a galaxy
and its parent cluster. If the cluster paradigm
held, there should be no trend. On the other hand,
if free expansion is a better model, galaxies with redshifts
smaller than the cluster mean are closer than those with
redshifts larger than the cluster mean. This distance
correlation translates into a trend of TF residuals
with relative redshift given by $\delta\eta \approx 5e
\log\left(v/v_{{\rm clust}}\right),$ where $e\approx 0.12$
is the inverse TF slope. This expected trend is indicated
as a dashed line in Figure~\ref{fig:etr_cp}. The trend in
the individual residuals is difficult to discern by eye because
of the large TF scatter. However, a running median of the
residuals, shown as connected solid squares, closely follows
the residual trend. This confirms that the large improvement
in the fit when the cluster paradigm is abandoned does in fact
derive from expansion motions of sample galaxies. Like the
samples of Bernstein \etal\ (1994) and Willick \etal\ (1995),
TF cluster samples in the LP10K ETR appear to be clusters in
name only; dynamically they more closely approximate field galaxies.

Thus, for the remainder of this paper we assume that the Hubble expansion
plus bulk flow model, equation~1, is the most accurate representation
of the redshift-distance relation for the LP10K galaxies. 
We note, however,
that the bulk flow derived from the cluster paradigm fit, line 4
in Table 2, differs by less than \onesigma\ in amplitude and
direction from the bulk flow derived with the expansion model.

\subsection{Main results}
\begin{table*}
\begin{minipage}{180mm}
\caption[ ]{\centerline{{\footnotesize
MAXIMUM LIKELIHOOD FIT RESULTS}}}
\centerline{
\begin{tabular}{crrrrclc}
\hline\hline \multicolumn{7}{c}{} \\ 
 &\multicolumn{1}{c}{$V_B^{(1)}$}&\multicolumn{1}{c}{$l$}&\multicolumn{1}{c}{$b$}&
\multicolumn{1}{c}{$\call$}&
$\chi^2_{{\rm clust}}$&Extinctions&Notes \\ \hline
\multicolumn{6}{l}{{\normalsize Results for ETR:}} & & \\
 &$ 961$&$266$&$19$&$-892.6$&$12.21$&(BH+SFD)/2 & 2,4 \\ 
 &$ 994$&$275$&$25$&$-891.5$&$13.94$&BH         & 2,4 \\ 
 &$ 970$&$252$&$12$&$-891.8$&$12.04$&SFD        & 2,4 \\ 
 &$ 805$&$252$&$18$&$-881.6$&$19.48$&(BH+SFD)/2 & 3,4 \\ 
 &$ 916$&$276$&$30$&$-892.0$&$11.60$&(BH+SFD)/2 & 2,5 \\ 
 &$ 202$&$343$&$52$&$-888.3$&$16.63$&(BH+SFD)/2 & 2,6 \\ 
 &$   0$&--&--&$     -888.0$&$17.92$&(BH+SFD)/2 & 2,7 \\  \hline 
\multicolumn{6}{l}{{\normalsize Results for FS:}} & & \\
 &$ 873$&$272$&$10$&$-1226.2$&$16.94$&(BH+SFD)/2 & 2,4 \\ 
 &$ 896$&$285$&$14$&$-1226.5$&$15.84$&BH         & 2,4 \\ 
 &$ 838$&$259$&$ 0$&$-1223.5$&$20.17$&SFD        & 2,4 \\ 
 &$ 766$&$276$&$30$&$-1226.1$&$17.79$&(BH+SFD)/2 & 2,5 \\ 
 &$ 151$&$343$&$52$&$-1221.7$&$24.81$&(BH+SFD)/2 & 2,6 \\ 
 &$   0$&--&--&$     -1221.4$&$25.72$&(BH+SFD)/2 & 2,7 \\ 
\hline\hline
\end{tabular}
}
\smallskip

Notes: (1) Best fit bulk flow amplitude, in \kms. This
value has {\em not\/} been corrected for the statistical
bias effect discussed in \S~4, and thus is \sm 25--30\% 
larger than the values quoted in the abstract, which
have been corrected.
(2) Expansion model used for redshift-distance relation 
(see text for details). (3) ``Cluster paradigm'' model
used for redshift-distance relation (see text for details).
(4) All three Cartesian components of bulk flow
vector varied in fit. (5) Bulk flow amplitude varied (negative
values allowed),
but direction fixed along CMB axis, $l=276\degs,$ $b=30\degs.$
(6) Bulk flow amplitude varied (negative
values allowed), but direction fixed along LP94 flow axis,
$l=343\degs,$ $b=52\degs.$
(7) Pure Hubble flow ($V_B \equiv 0$) assumed.
\end{minipage}
\end{table*}

The main results obtained from the maximum-likelihood fits are
presented in Table~2. The bulk flow parameters are expressed in
terms of amplitude and direction in columns 1--3. The amplitudes
have {\em not\/} been corrected for the bias
discussed in \S~5. The directions are unbiased, however (\S~5).
Columns (4) and (5) list the likelihood and $\chi^2_{{\rm clust}}$ statistics
defined in \S~3. Column (6) gives the extinction method used in each fit;
for reasons discussed in \S~4.1, we now focus only on fits using the
mean extinctions, (BH+SFD)/2. Column (7) provides keys to details
of the fits given in the table notes.

Our final bulk flow vectors are obtained from the likelihood
fits in which all three components of the flow velocity
were varied, namely, the first lines in the ETR and FS sections
of the table. The flow amplitudes, $961\ \kms$ for the
ETR and $873\ \kms$ for the FS, are biased high by 
33\% and 25\% respectively, as determined from the Monte Carlo
analysis of \S~5. When corrected for these biases, they yield
the flow amplitudes quoted in the Abstract and Summary, \S~6.3.
The cluster-$\chi^2$ statistic
does not necessarily take on its minimum value for these fits, 
because the flow parameters are computed by maximizing likelihood,
not by minimizing $\chi^2_{{\rm clust}}.$ However, the
maximum-likelihood fits produce values of $\chi^2_{{\em clust}}$
within \sm 1 unit of its minimum value. 
 
Other entries in Table~2 test alternative flow directions. The lines
in which the direction is given as $l=276\degs,$ $b=30\degs$
give results of fits in which the flow 
was assumed \apriori\ to be parallel to the LG peculiar velocity,
and only the amplitude was varied. The likelihood statistics
for these fits differ very little from the best-fit values. Indeed, this choice of
flow direction produces a smaller $\chi^2_{{\rm clust}}$ than 
does the best fit for the ETR. The LP10K bulk
flow may thus be described, to better than \onesigma\ accuracy, as being
in the same direction as the LG motion. 

The lines in Table~2 in which the direction is given as
$l=343\degs,$ $b=52\degs$ give results of fits in which the flow
was assumed \apriori\ to be parallel to the LP94
bulk flow. The best-fit flow amplitudes along this
axis are much smaller, for both the ETR and FS, than the
700 \kms\ reported by LP94. Moreover, the likelihood and $\chi^2_{{\rm clust}}$
values for this fit indicates that these solutions are poor models
in comparison with that in which the flow is held parallel to the
LG motion. As these two models have the same number of free parameters,
these differences are highly significant. In short, the LP10K data set
is inconsistent with the LP94 bulk flow.

Table~2 also lists the fit results when $V_B\equiv 0,$ i.e., pure
Hubble flow in the CMB frame. This yields the lowest likelihood 
and largest $\chi^2_{{\rm clust}}$ values of all fits considered (except
the cluster paradigm fit discussed in \S~4.2). We defer to \S~5 a quantitative
discussion of the confidence level with which the data set enables
us to rule out the $V_B\equiv 0$ model. However, from the change
in the cluster-$\chi^2$ statistic between the no-flow
model and the best-fit model for the FS,
$\Delta\chi^2_{{\rm clust}}=8.78$ with the addition of
three degrees of freedom, we can estimate that the
$V_B\equiv 0$ model is ruled out at about the $2.4\sigma$ level.
Our Monte-Carlo assessment of significance levels (\S~5) 
roughly confirm this estimate.

In Figures 3 and 4 we plot CMB frame radial peculiar velocities of
the 15 LP10K clusters, for the ETR and FS samples respectively.
The velocities plotted are those that went into the $\chi^2_{{\rm clust}}$
calculations for the $V_B=0$ model. They are derived from the
mean inverse TF residuals for each cluster, assuming that all
cluster galaxies lie at the mean cluster velocity. (In the case of
the FS, this is not always a good assumption, but it is the only way to
assign a peculiar velocity to a given cluster.) The point sizes are
inversely proportional to peculiar velocity errors, so that
large symbols carry proportionally more weight in determining
the bulk flow solution. In each figure error bars are drawn
for two points to calibrate the size-error mapping. In Figure 3,
the velocities are plotted against the cosine of the angle
between the cluster and the CMB dipole
direction ($l=276\degs,$ $b=30\degs$), the flow direction
which yielded the smallest $\chi^2_{{\rm clust}}$ among the
ETR fits. In Figure 4, the velocities are plotted against the
cosine of the angle between the cluster and $l=272\degs,$
$b=10\degs,$ the flow direction which
yielded the smallest $\chi^2_{{\rm clust}}$ for the FS.
The tilted solid lines in each figure represent the predicted radial
peculiar velocities from the respective flow models.

Figures 3 and 4  provide visual evidence of the bulk flow
solutions found from the likelihood analysis. In each case, the
clusters near the apex of of the flow axis have peculiar velocities
which are positive in the mean; clusters nearer the antapex tend to
have negative velocities. The notable exceptions to the trend
are the clusters A2657 and A260, which lie near the antapex of the
flow and have velocities $\simgt 0.$ However, as indicated
on the figures, these clusters have comparatively few members.
A2657 is in consistent at the $<1\,\sigma$ level with the
flow model for the ETR, and A260 is consistent with the
flow model for the FS. Thus, although these clusters deviate
from the trend, the deviation is not statistically significant.
For the FS, the cluster A2247 is a large ($\sim 2\,\sigma$) outlier
from the flow model, but in a sense which reinforces the flow.

While Figures 3 and 4 show why the data favor a
large bulk flow in the general direction of the CMB dipole,
they also reflect the relatively large scatter in the
TF relation. At the distances of the LP10K clusters, even
samples of 10--20 TF galaxies give rise to \onesigma\
peculiar velocity errors approaching 1000 \kms. As a result,
there is considerable scatter about the mean trend in the diagrams,
and the flow solutions themselves have
significant errors. 

\vbox{%
\begin{center}
\leavevmode
\hbox{%
\epsfxsize=8.9cm
\epsffile{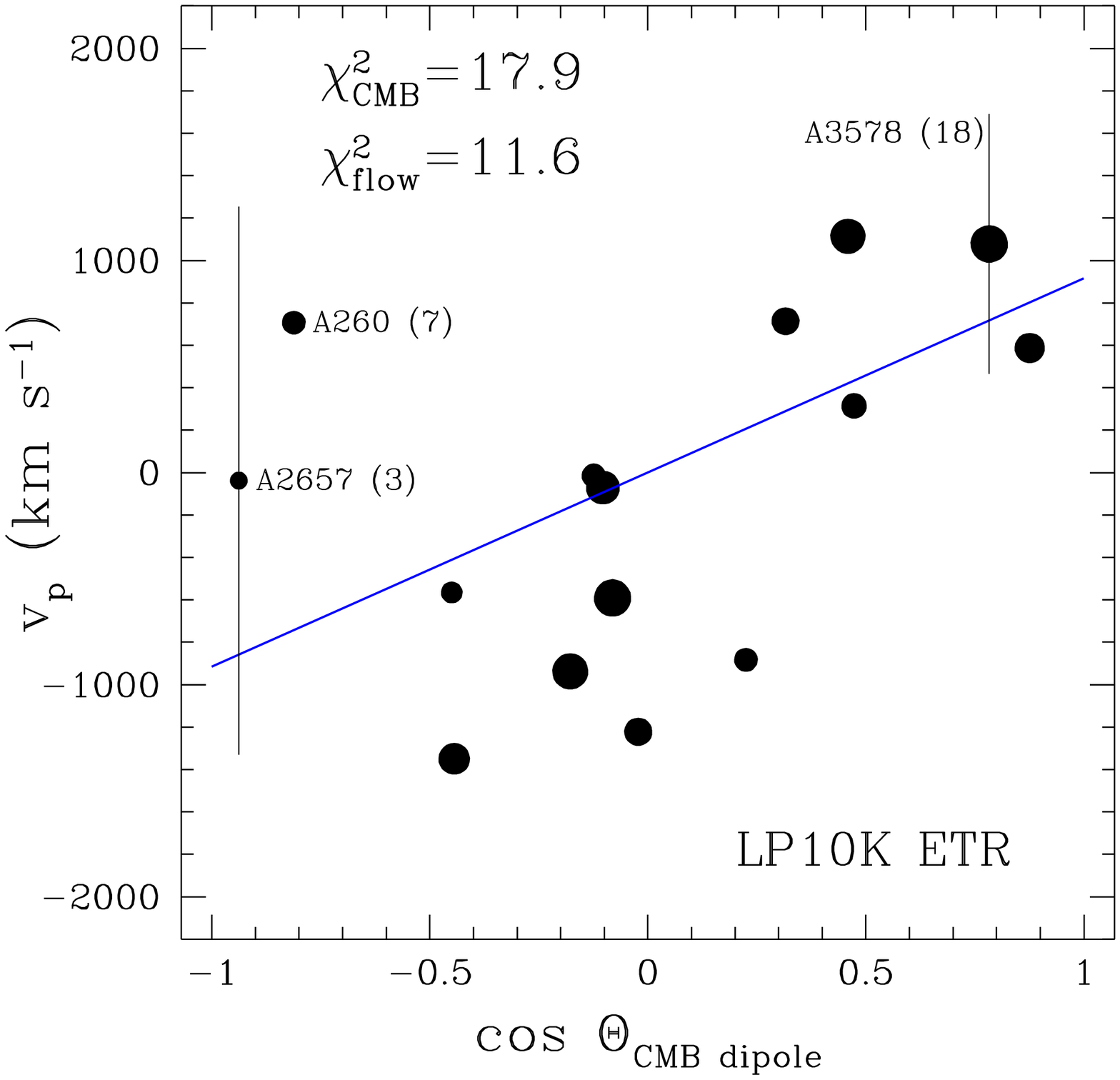}}
\begin{small}
\figcaption{%
 CMB frame peculiar velocities of the 15 LP10K clusters, plotted
against the cosine of the angle between the cluster position
on the sky and the apex of the CMB dipole, $l=276\degs,$
$b=30\degs.$ The cluster TF
samples are restricted to the ETR.
Point sizes
are inversely proportional to the peculiar velocity errors.
The largest (A2657) and smallest (A3578) of these
errors are shown as error bars, with the number of ETR
galaxies in those clusters also indicated. The cluster
which deviates most significantly from the bulk flow
fit, A260, is also labeled.
The tilted line shows the best-fit bulk flow
amplitude, $916\ \kms,$ when the 
when the bulk flow direction is fixed toward the CMB dipole.
See text for further details.
\label{fig:clust_etr}}
\end{small}
\end{center}}

\vbox{%
\begin{center}
\leavevmode
\hbox{%
\epsfxsize=8.9cm
\epsffile{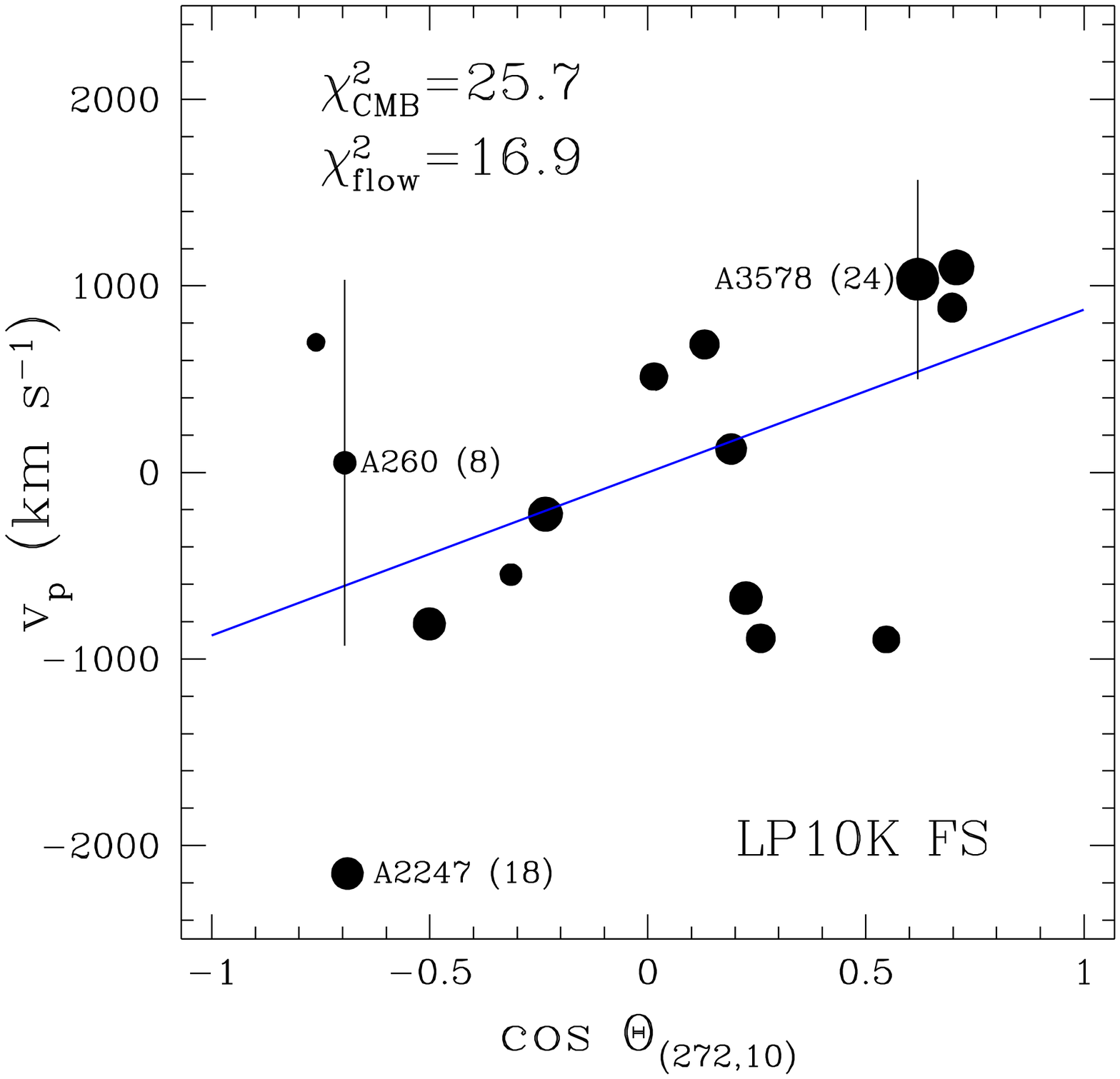}}
\begin{small}
\figcaption{%
Same as the previous figure, except 
cluster peculiar velocities
for the LP10K {\em full sample\/} are plotted 
now against the cosine of the angle between the cluster position
on the sky and best-fit flow direction for the FS,
$l=272\degs,$ $b=10\degs.$ Peculiar velocity error bars
are indicated for A3578 and A260. Also identified is
A2247, the only $2\,\sigma$ outlier from the fit.
\label{fig:clust_fs}}
\end{small}
\end{center}}

\subsection{TF parameters}
\begin{table*}
\begin{minipage}{180mm}
\caption[ ]{\centerline{{\footnotesize
TULLY-FISHER PARAMETERS}}}
\centerline{
\begin{tabular}{c c c c c c c | l} \\ \hline\hline 
\multicolumn{8}{c}{} \\
$D$ & $e$ & $\alpha$ & $\beta$ & $f_s$ & $\sigI$ & $\delta v_{rot}$ &
FS/ETR \\ \hline
$-21.664$&$0.1138$&$0.0485$ &$0.0427$&$1.993$&$0.0227$ & $17.5$ & ETR \\ 
$-21.675$&$0.1170$&$0.0500$ &$0.0407$&$2.029$&$0.0302$ & $17.7$ & FS \\ 
\hline\hline
\end{tabular}
}
\smallskip

Notes: TF parameters obtained from the best-fit maximum likelihood
flow solutions in Table~2, for both the ETR and the FS. The meaning
of the various parameters is given in \S~3. The quantity $\delta v_{rot}$
is expressed in \kms.
\end{minipage}
\end{table*}

In Paper I we applied a pure Hubble flow model to the LP10K
full sample in order to study the properties of the TF
relation. Here we have shown that the addition
of a bulk flow to the redshift-distance model significantly
improves the fit likelihood. It is thus worth asking
whether the TF parameters themselves have changed significantly.
The TF parameters derived from the best-fit ETR and FS flow
models are listed in Table~3. The meaning of the parameters
given in the table was described
briefly in \S~3, and in greater detail in Paper I.

Table~3 should be compared with Table~2 (line 4) of Paper I, which lists
the same parameters from the Hubble flow fit of that paper.
The parameters are seen to have changed very little as
a result of the flow model and the adoption of a more
rigorous likelihood algorithm. In particular, the values
of $\alpha,$ which measures
the surface-brightness dependence of the
TF relation, and of $f_s,$ the number of disk
scale lengths at which the rotation curve should
be evaluated to optimize the TF relation, and to which
particular physical significance was ascribed
in Paper I, are seen
to be virtually unchanged. The TF slope and zero point
have changed very slightly as a result of adopting
the flow model.

A somewhat more significant change is seen in the value
of the rotation velocity measurement error $\delta v_{rot}$ (called
$\delta \vtf$ in Paper I), which has increased by about 10\%. 
This change may be ascribed, as discussed in \S~3, to
the proper accounting for repeat spectroscopic measurements
and photometric errors
in the improved likelihood method of this paper.
As a result, $\delta v_{rot}$ is more accurately
determined. This, in turn, means that the intrinsic
TF scatter 
is better constrained than in Paper I. The value of $\sigI$ for
the FS, $0.030,$ is the more reliable because of the
larger sample size. The Monte-Carlo simulations enable
us to test for biases in the TF parameters just as
the bulk flow parameters, and they show that
the maximum likelihood value of $\sigI$ is biased
low by about 10\%. They also show that the rms error
in the determination of $\sigI$ is about 25\%. Taking
these into account, out best estimate of the inverse
TF scatter is $\sigI=0.033\pm 0.008$ dex. 
The corresponding value of the
forward TF scatter is $\sigma_{I,forw} = \sigI/e = 0.28\pm 0.07$ mag.

\section{Monte-Carlo Simulations}
\subsection{Details of the simulations}
The true sky positions, redshifts, apparent magnitudes,
and inclinations of all LP10K TF galaxies were used as initial input to
the simulations. Distances were assigned to each galaxy according
to a bulk flow model, $d=H_0^{-1} [cz-\bfv_p\cdot\nhat].$ 
Absolute magnitudes, $M=m-5\log(d)-25,$
where thus derived, and a preliminary circular velocity parameter
$\eta_0=-e(M-D)$ was assigned to each galaxy.
The TF
parameters were taken to be $e=0.12$ and $D=-21.62,$
similar to the observed values. The surface-brightness and concentration-index
dependences of the TF relation were neglected in the simulations.

A Gaussian random variable of mean zero and dispersion $\sigma_I=0.032$
was added to $\eta_0,$ yielding a ``true'' circular velocity
parameter $\eta_1$ and a corresponding projected rotation
velocity $v_{rot}^{proj}=158.1 \sin i \times 10^{\eta_1}\ \kms,$ 
where $i$ was the observed galaxy inclination. A second
Gaussian random variable of
dispersion $16\ \kms$ was added to $v_{rot}^{proj}$
to simulate the effect of
spectroscopic measurement errors. If there were two spectroscopic
observations of an individual galaxy, independent random errors
were added to each.
The scattered rotation velocities were then divided
by $\sin i$ to produce the final observational values of $v_{rot}$
and $\eta.$ 
Finally, Hubble flow
noise was simulated by scattering each redshift by a Gaussian
random variable of dispersion $\sigv=250\ \kms.$ 

The above procedure
yields a simulated data set that mimics in most respects the statistical
properties of the real one. Photometric errors were not
included in the simulations, but these have a negligible
effect on the overall TF scatter.
The simulated data sets were then subjected to the same
cuts on absolute magnitude, inclination, etc.\ (cf.\ Paper I) as the real one.
Consequently, the results of applying the maximum likelihood code to them
should faithfully reflect the random errors in the recovered
peculiar velocity vector.

\subsection{Choice of simulated bulk flow vectors}

Ideally one would like to simulate a suite of universes with
a large variety of underlying bulk flow vectors. One could then
ask, for any given flow vector, what is the probability of
obtaining the result found from the real data? One could then 
invert this relation using Bayesian statistics and
obtain the probability distribution of the true vector given
the observed data. 
However, because of the relatively large observational errors for LP10K, 
it is not essential to carry out this full-blown analysis here.
For our present purposes, two types of simulations, each corresponding
to a reasonable ``paradigm,'' suffice:
\begin{enumerate}
\item Paradigm I. The Hubble flow beyond \sm 10,000 \kms\ is isotropic
in the CMB frame, i.e., $V_B=0.$ The
only perturbation to Hubble expansion are random
velocities. 
\item Paradigm II. The motion of the Local Group relative to the 
CMB is generated on very large scales,
$\simgt 300\h1$ Mpc.
The bulk flow of the LP10K sample would then be
similar to the peculiar velocity of the Local Group. For
these simulations we adopt
$V_B=625\ \kms$ toward $l=270\degs,$ $b=30\degs.$ 
The random velocities are, of course, still present.
\end{enumerate}

\subsection{Results of simulations}
Four Monte-Carlo runs were carried out in total, two based
on paradigm I and two on paradigm II. For each paradigm the
FS and ETR
samples were simulated. Each of
the four runs consisted of $10^4$ simulated data sets
and likelihood anlayses. The results
of each simulation were a recovered bulk flow components
$V_x,$ $V_y,$ and $V_z,$
as well as the various TF parameters and likelihoods.
\vbox{%
\begin{center}
\leavevmode
\hbox{%
\epsfxsize=8.9cm
\epsffile{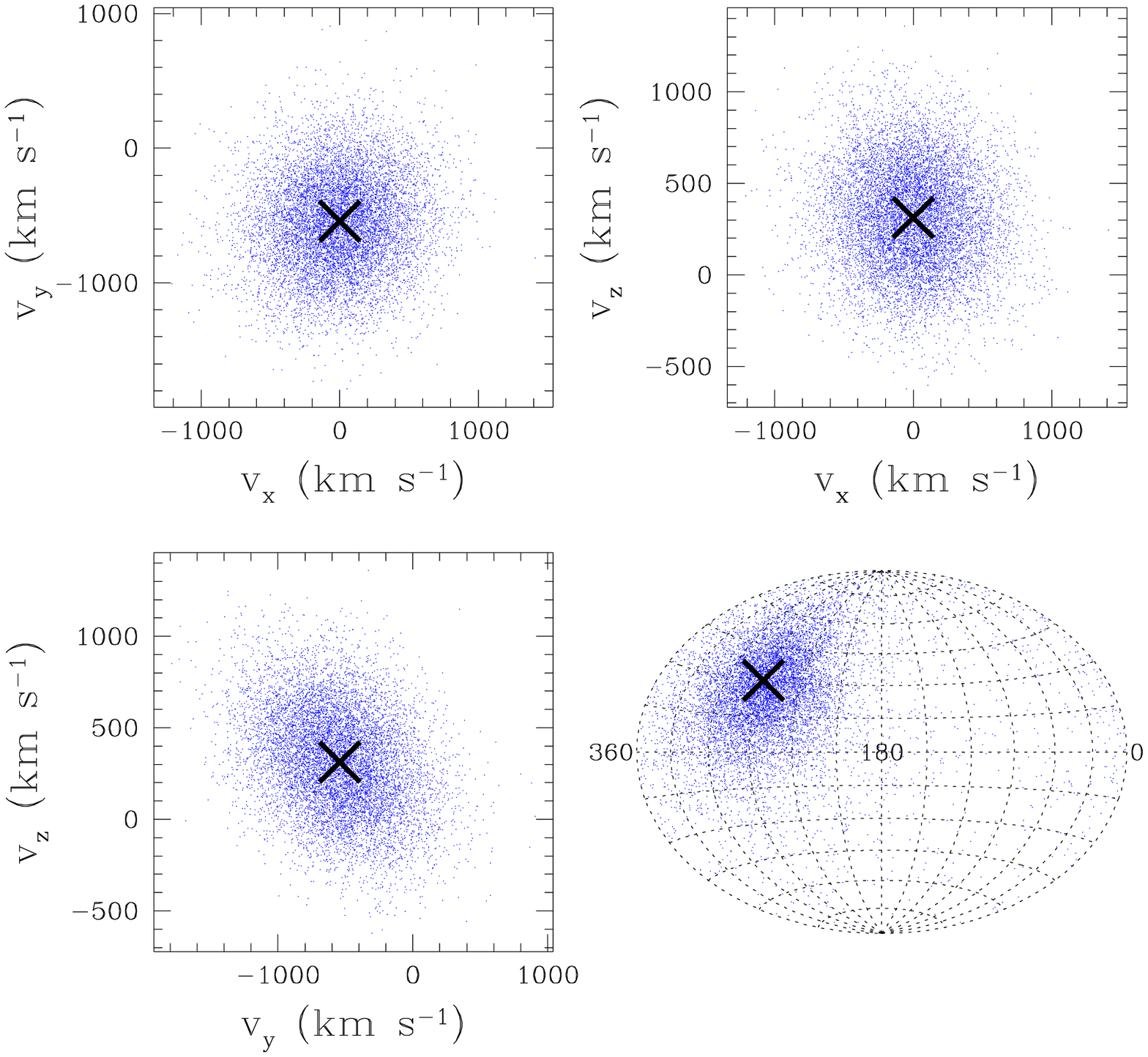}}
\begin{small}
\figcaption{%
Ten thousand random realizations of the LP10K full sample in
a universe in which the true bulk flow is $625\ \kms$
towards $l=270\degs,$ $b=30\degs.$ The recovered Galactic Cartesian
velocity components are plotted in the upper panels and
the lower left panel; the lower right panel shows
the recovered direction of the flow
in Galactic coordinates.
\label{fig:sim_f6}}
\end{small}
\end{center}}

Figure~\ref{fig:sim_f6} shows the results of simulating
the LP10K full sample based on paradigm II. (The plot that results
from simulations of the ETR subsample is quite similar
in appearance.) Several key features of the plot may
be noted. First, the recovered individual velocity components
are very nearly unbiased. The input components of the
bulk flow are $V_x=0,$ $V_y=-541.3,$ and $V_z=312.5\ \kms.$
The corresponding mean recovered values are
$\vev{V_x}=-7.8\pm 3.2,$ $\vev{V_y}=-554.0\pm 3.5,$ and
$\vev{V_z}=318\pm 2.8\ \kms.$ 
Thus, the individual components are only very slightly 
($\simlt 15\ \kms$) biased. This bias is very small in 
comparison with the rms scatter in the derived
components, which is \sm 300 \kms. The accuracy
of the components similarly translates into a
mean flow direction toward $l=270.3\degs,$ $b=30.1\degs,$
virtually identical to the input direction. The rms error
in the flow direction about the mean is 35\degs. Thus, 
the simulations demonstrate that {\em the LP10K can recover
the direction of the bulk flow, and the value of its individual
Cartesian components, in an unbiased fashion.} The residual
biases are so small in comparison with the scatter of
a single measurement that they may be neglected.

The amplitude of the recovered flow vector is, on the other hand,
biased high. This occurs because the scatter in the individual
components can only increase their quadrature sum. To quantify
this bias we take the average of all $10^4$ recovered
velocity amplitudes, which we find to be $784\ \kms,$ or
$25.4\%$ higher than the input value of $625\ \kms.$
We take this as a measure of the velocity amplitude
bias for the LP10K FS. For the ETR, the bias is somwhat larger, $33.1\%$
Thus, the final quoted values for the bulk flow derived from the
FS and ETR samples are corrected by factors $(1.254)^{-1}$ and
$(1.331)^{-1},$ respectively, relative to the values which appear
in Table~2.

We calculate the error in the
recovered bulk flow amplitude as the average over $10^4$ simulations
of $|V_B - \vev{V_B}|.$ For FS, this yields $\Delta V = 250\ \kms.$
For the ETR the same procedure yields $\Delta V = 280\ \kms.$ These
are the estimates of the \onesigma\ bulk flow amplitude errors
given in the absract. 

\vbox{%
\begin{center}
\leavevmode
\hbox{%
\epsfxsize=8.9cm
\epsffile{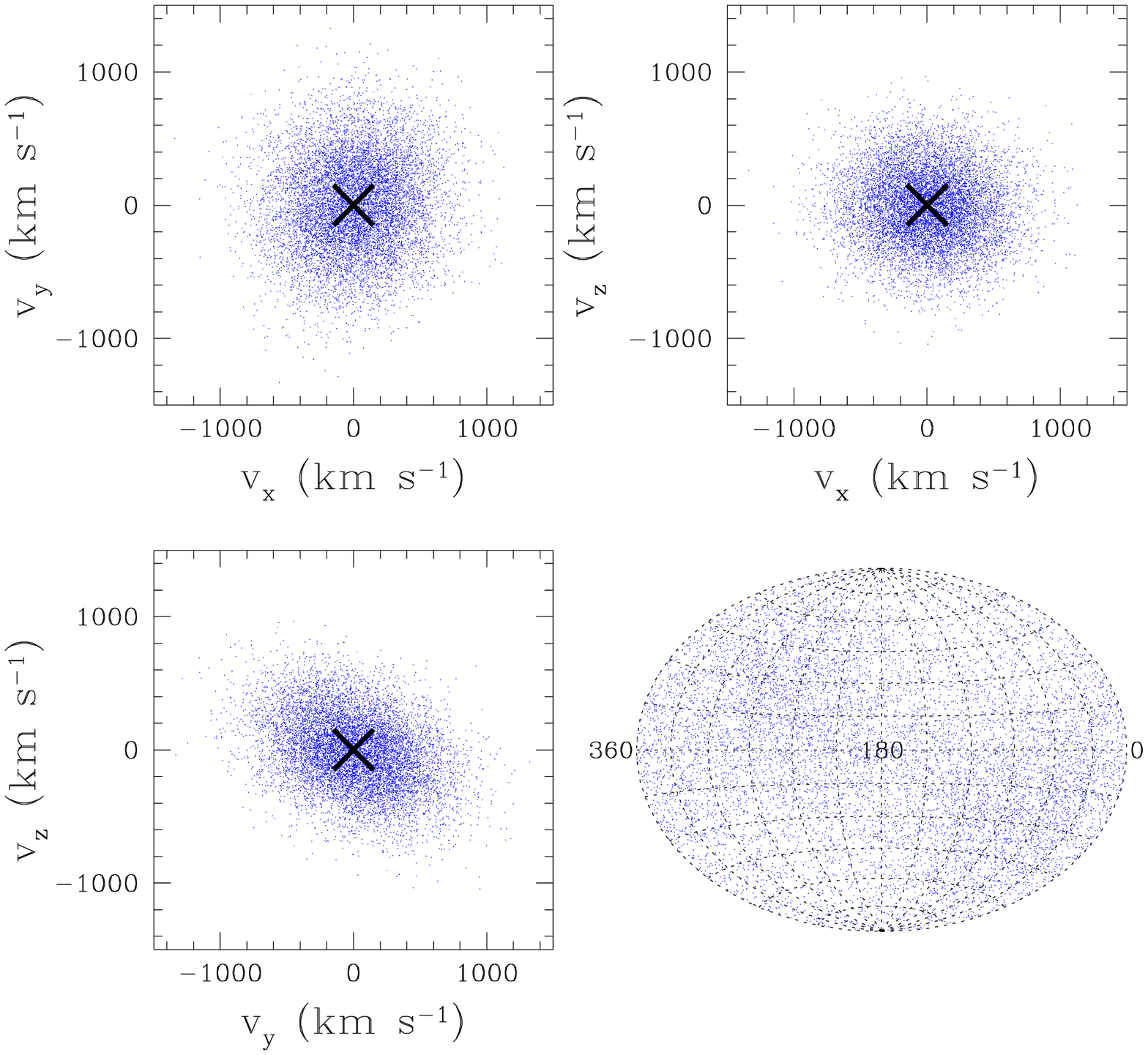}}
\begin{small}
\figcaption{%
Ten thousand random realizations of the full LP10K sample in
a universe with zero bulk flow in the CMB frame.
The quantities plotted are the same as in the previous figure.
\label{fig:sim_f0}}
\end{small}
\end{center}}


The paradigm II simulation enabled us to estimate biases and errors
under the assumption that the detected ETR and FS flows
are real. In order to measure how well the LP10K data set rules out
the hypothesis of convergence, however, we need to consider the
paradigm I simulation, in which the true bulk flow vanishes
in the CMB frame. The results of $10^4$ paradigm I simulations
of the LP10K FS sample are shown in Figure~\ref{fig:sim_f0}. As before,
heavy crosses mark the true values of the Cartesian components of
the flow, in this case, $V_x=V_y=V_z=0.$ 
The average values recovered from the simulations are $V_x=3.8\pm 3.2\ \kms,$
$V_y=5.2\pm 3.5\ \kms,$ $V_z=-12.6\pm 2.7\ \kms.$ 
Thus, the likelihood analysis again recovers,
on average, the true flow components with negligible bias.
The lower right panel of Figure~\ref{fig:sim_f0} demonstrates
that there is no preferred direction for the recovered velocity
vector. The points are well-distributed around the sky, although
a slight preference for the CMB apex and antapex quadrants is
apparent. (This tendency is also manifested in the non-circular
shape of the $V_y$ versus $V_z$ plot.) This represents a small
geometrical bias that results from the imperfect isotropy
of the LP10K clusters.

To determine the confidence level at which we can rule out convergence,
we proceed as follows. We ask the question, what fraction of paradigm I
simulations yield derived flow vectors of amplitude $V_B \ge f V_{data},$
and in a direction $\nhat$ such that $\nhat\cdot\nhat_{data} \ge 
\cos\theta_{{\rm RMS}},$ where $\theta_{{\rm RMS}}$ is the \onesigma\
directional error of the flow, and $V_{data}$ and $\nhat_{data}$ are
the derived amplitude and direction of the LP10K observed (ETR and FS) flows.
We choose the parameter $f$ such that {\em half\/} of all
paradigm II simulations satisfy these threshold
criteria, with $V_{data}$ and $\nhat_{data}$ are replaced by the
mean paradigm II amplitude and direction; this yields $f\approx 0.8.$
In other words, we ask, ``What fraction of
zero-flow simulations produce flow results that occur half
the time when the flow is real?'' The answer to this
question gives the probability that the LP10K results occur
by chance in a $V_B=0$ universe.

When we carry out this exercise, we find that 5.3\% (525/10000) of
all ETR paradigm I simulations, and 2.9\% (290/10000) of the FS
paradigm I simulations, meet these criteria.
We thus deduce that we can rule out the hypothesis
that the Hubble flow has converged to the CMB frame at distances
less than \sm 100\h1\ Mpc at the 94.7\% confidence level from
the ETR subsample. Using the full LP10K TF sample, we can
rule out the convergence hypothesis at the 97.1\% confidence level.
These significance levels are roughly consistent with
those we deduced in \S~4.3 from changes
in the $\chi^2_{{\rm clust}}$ statistic between the no-flow
and best-fit models in Table~2.

\section{Further Discussion and Summary}
The results presented in this paper suggest that the volume
of the local universe within \sm 15,000 \kms\ of the LG
possesses a bulk peculiar velocity of \sm 450--950 \kms\ with
respect to the CMB frame. 
In this section, we discuss the 
significance of this finding.

\subsection{How reliable are the results?}
first, we must ask, is the LP10K bulk flow real? There are several
reasons for caution. The statistical significance level is not high,
somewhat greater than $2\,\sigma.$ 
A result at such a modest significance level must
be confirmed with independent data sets. We reiterate that the
LP10K survey includes elliptical galaxy FP data in the same fifteen
clusters. In several years the observations and reduction of the elliptical
sample will be completed, and will provide an independent check of
the flow measured from spiral galaxies using the TF relation.
 
There are other recent flow analyses to compare with the present
result. The most encouraging comparison is with the recently
completed SMAC survey of Hudson \etal\ (1998a,b). 
The SMAC sample is about 75\%
as deep as LP10K, but has similarly good sky coverage,
and involves many more galaxy clusters and individual
galaxies. Moreover, SMAC is an FP survey of elliptical
galaxies, and thus is completely independent from the LP10K TF data.
It is thus noteworthy that the bulk flow vectors obtained from the
two programs are in good agreement in both amplitude and
direction (cf.\ \S 1). The two results together constitute a strong argument
for the reality of the flow.

The LP10K and SMAC findings are, however, at variance with
the results of the SFI/SCI survey of Giovanelli and coworkers
(Giovanelli \etal\ 1998a,b; Dale \etal\ 1998). SFI/SCI
is an $I$-band TF survey comprising numerous cluster and
field galaxies. Its depth is somewhat less than LP10K's,
but quite similar to SMAC's, at 8000--10,000 \kms. The
most consistent result from SFI/SCI is the clear signal
of convergence of the Hubble flow to the CMB frame by
a distance of 6000 \kms. At 10,000 \kms, the SFI/SCI
group finds the bulk flow to be $\le 200\ \kms$ with
high confidence, and to be consistent with zero (Dale \etal\ 1998).

There is no simple way at present to resolve the discrepancy
between the LP10K and SMAC results, on the one hand,
and those of the SFI/SCI project, on the other. What will be
required is a systematic comparative analysis of the data
sets on a cluster by cluster, and object by object, basis.
Such a comparison will be possible when all three groups
have published their complete
data sets. This will occur in the near future for the
LP10K TF sample in Paper III of this series.

\subsection{Interpretation}

The discussion above shows that the evidence for a $\simgt 600\ \kms$
bulk flow on a $\simgt 150\h1$ Mpc scale is suggestive
but not yet compelling. 
Bearing this in mind, let us for the moment
take the LP10K and SMAC results at face value and ask,
what are the implications for cosmology? Peculiar velocities
arise naturally within the gravitational instability scenario
for structure formation. 
The amplitude and
coherence scale of bulk flows thus reflect both the 
power spectrum of density fluctuations and the value
of the density parameter $\Omega_M,$ and
can, in principle, provide constraints on these important cosmological quantities.

\def\wt{\widetilde}
More specifically, consider 
the mean square amplitude of the bulk velocity
on a scale $R.$ 
According to linear gravitational instability theory
it is given by (e.g., Strauss \& Willick 1995)
\begin{equation}
\vev{v^2(R)} = \frac{H_0^2\Omega_M^{1.2}}{2\pi^2} 
\int_0^\infty P(k)\, \wt W^2(kR) \, dk\,,
\label{eq:vofr}
\end{equation}
where $\Omega_M$ is the
present value of the matter density parameter, 
$P(k)$ is the mass fluctuation power spectrum, and $\wt W(kR)$ is
the Fourier transform of the window function of
the sample, $W(r).$ 
Precise determination of the
window function is notoriously difficult
for real samples. A common approximation,
the ``tophat'' window function, clearly
does not apply to the LP10K sample, which attempted
to sample a shell rather than a full spherical volume.
When restricted to the extended target range,
LP10K does indeed approximate a shell, with
selection probability roughly constant for
between $R_1=90\h1$ Mpc and $R_2=130\h1$ Mpc.
A reasonable representation for the ETR window function is thus
\begin{equation}
W_{{\rm ETR}}(r) = \frac{3}{4\pi(R_2^3-R_1^3)} \times 
 \left\{ \begin{array}{ll}
1\,, & R_1 \leq r \leq R_2 \,;\\
0 & {\rm otherwise}\,. \end{array} \right.
\label{eq:winlp10k}
\end{equation}
The corresponding window function in Fourier
space is
\begin{equation}
\wt W_{{\rm ETR}}(x_1,x_2) = \frac{3}{f^3-1}\left[f^3\frac{j_1(x_2)}{x_2}
- \frac{j_1(x_1)}{x_1}\right]\,,
\label{eq:winetr}
\end{equation}
where $x_1=kR_1,$ $x_2=kR_2,$ $f=R_2/R_1,$ and $j_1(x)$
is the first spherical Bessel function. (The reader
may note that this form of the window function tends
to the usual tophat form in the limit $R_2\gg R_1.$)
Substitution of equation~\ref{eq:winetr}
into equation~\ref{eq:vofr} yields, for an adopted 
cosmology and power spectrum, the mean square amplitude
of the bulk flow for the LP10K ETR subsample.

\def\vrms{V_{{\rm RMS}}}
Before making this calculation, a related issue should be clarified.\footnote{The
author is indebted to Paul Steinhardt
for insightful comments on which the following discussion is based.}
The value of $\vev{v^2}$ in a given window is three times the mean square value
of any individual Cartesian velocity component. Consequently, the 
velocity amplitude
has a Maxwellian distribution, $P(v) dv \propto v^2 \exp(-3v^2/2\vev{v^2}).$
The significance of the observed velocity must be
gauged relative to this distribution. In particular, one may choose a significance
level $F,$ say, and a corresponding velocity
amplitude $v_F,$ such that $P(v\ge v_F) = 1-F.$  A brief calculation then
shows that 
\begin{equation}
v_F = \sqrt{\frac{2}{3}} \vrms\,y_F\,,
\label{eq:vfyf}
\end{equation}
where $\vrms = \sqrt{\vev{v^2}},$ and $y_F$ is the solution to the equation
\begin{equation}
F = \erf (y_F) - \frac{2 y_F}{\sqrt{\pi}} e^{-y_F^2}\,.
\label{eq:fyf}
\end{equation}
With this procedure we find, for example, that 1\% of volumes
will exhibit $v\ge 1.945\,\vrms$ ($F=0.99$), while 0.1\%
will exhibit $v \ge 2.329\,\vrms$ ($F=0.999$). (This calculation
refers only to cosmic variance, and neglects observational error.)

\vbox{%
\begin{center}
\leavevmode
\hbox{%
\epsfxsize=8.9cm
\epsffile{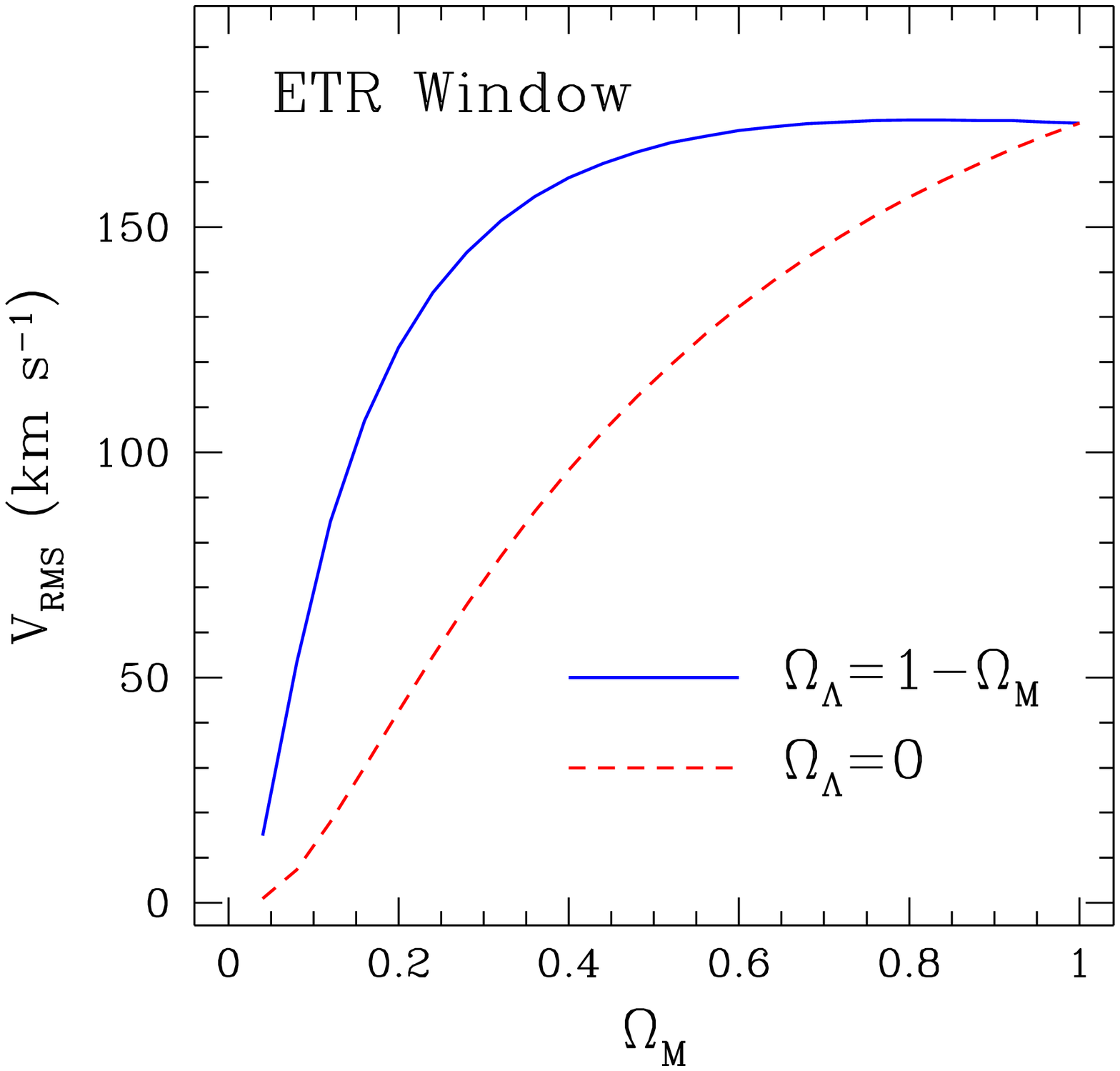}}
\begin{small}
\figcaption{%
The predicted RMS velocity for a shell of inner radius
90\h1\ Mpc and 130\h1\ Mpc, the volume probed by
the LP10K ETR subsample. Scale-invariant, 
COBE-normalized power spectra are used in the calculations.
The solid line shows predictions for a spatially flat
universe, the dashed line for a $\Lambda=0$ universe.
\label{fig:vrms_etr}}
\end{small}
\end{center}}

Figure~\ref{fig:vrms_etr} shows the results of calculating $\vrms$
through the LP10K ETR window, for $0 < \Omega_M \leq 1$
and COBE-normalized CDM-type power spectra. A Hubble
constant of $65\ \kmsmpc$ and a baryon abundance $\Omega_b h^2 =0.025$
were adopted for the calculation. 
It is evident
that $\vrms$ is quite small in comparison with
the measured value of $720\ \kms$ for the LP10K ETR. If, for
example, we consider cosmological parameters favored by
recent measurements of Type Ia Supernovae, $\Omega_M=0.3,$
$\Omega_\Lambda=0.7$ (Riess \etal\ 1998; Perlmutter \etal\ 1999),
$\vrms=150\ \kms.$ Thus, 99\% of ETR volumes in the universe
should exhibit bulk flows less than 292 \kms\ in such a univers,
and 99.9\% of such volumes should possess bulk flows less than 349 \kms.
The corresponding values for $\Omega_M \approx 1$ are only about 17\%
larger. For low-density, open universes $\vrms$
is several times smaller than
for a flat universe of the same $\Omega_M.$

We have also shown that the LP10K full sample, more than 25\%
of whose members have $0.05 < z \leq 0.1$  and thus lie well beyond
the ETR window, yields as strong a signal
of large-scale bulk flow as the ETR. It is more difficult
to calculate expected theoretical values for the FS, because its
window function is poorly defined. However, we 
gain insight into the effect of the added depth by crudely
approximating the FS window function as having the same
form as the ETR's, equation~\ref{eq:winlp10k}, but with
$R_2=180\h1$ Mpc. (We leave the
lower limit at 90\h1\ Mpc, as there are only 8 FS galaxies in
the foreground of the ETR.) 
When $\vrms$ is calculated for this window function, one finds
it to be \sm 20--30\% smaller,
for given values of $\Omega_M$ and $\Omega_\Lambda,$ 
than it was for the ETR window. 
Thus, the FS bulk flow of $700\ \kms,$
taken at value, exceeds theoretical predictions by a
larger margin than the ETR result.

A more thorough exploration of parameter
space could, of course, yield slightly larger values of $\vrms.$ This
is not necessary here, because the errors on the LP10K flow amplitudes
are too large to strongly rule out models at present. The important point
is simply that bulk flows on $\simgt 100\h1$ Mpc scales are
well-suited as probes of large-scale power. In particular,
the results of this paper, if ultimately proven correct, would indicate
the need for models with more power on large
scales than COBE-normalized CDM models can provide. A similar conclusion
was reached for the LP94 bulk flow
by Strauss \etal\ (1995).

We can anticipate future data sets that will probe
bulk streaming on scales $\simgt 300\h1$ Mpc. Such measurements will
not be possible with methods such as TF and FP, for
which peculiar velocity errors become prohibitively
large at $z\simgt 0.1.$ Type 1a supernovae
(e.g., Riess \etal\ 1997) are sufficiently accurate
distance indicators for this purpose, although it may prove difficult to assemble
the required full-sky samples. Perhaps the most promising 
approach is measurement of the Sunyaev-Zel'dovich (SZ)
effect in rich clusters. When combined with X-ray
data, SZ measurements yield peculiar velocity estimates
with an accuracy of 500--1000 \kms\ (Holzapfel \etal\ 1997).
Moreover, SZ measurement errors,
although relatively large, do not increase with distance (in
contrast to those of methods such as TF, FP and SN Ia),
and thus provide a unique probe of bulk flows on scales $z\simgt 0.1.$

It is notable that the
LP10K and SMAC flow vectors are within \sm 30\degs\
of the CMB dipole. If one subtracts the LG velocity from these
vectors, little or no motion is left.  Should the same  pattern
hold on much larger scales, one might legitimately question
the special character we assign the CMB frame of reference in
analyzing the Hubble expansion. Of course, to do so
would also require a non-kinematic explanation of the
$\Delta T/T \sim 10^{-3}$
dipole anisotropy which is 
consistent with the fact that higher-order anisotropies are
two orders of magnitude smaller. 
Such scenarios
have been proposed in the past (e.g., Paczynski \& Piran 1990), but they
require substantial modifications of standard Big Bang cosmology.
The present data do not compel us to seriously consider such
alternatives, but it is worth bearing in mind that future data
could push us in that direction.

\subsection{Summary}

We have used the LP10K TF sample to measure the bulk
peculiar velocity on a $\simgt 150\h1$ Mpc scale.
Both the full sample of 244 galaxies, and a subsample
restricted to galaxies with $7000 \le cz \le 15,\!000\ \kms,$
the ``extended target range'' or ETR, were considered. The
derived bulk flow was the same, within the errors, for
both samples: $v_B = 720\pm 280\ \kms$ toward
$l=266\degs,$ $b=19\degs$ for the ETR, and $v_B = 700\pm 250\ \kms$ toward
$l=272\degs,$ $b=10\degs$ for the full sample. 
The overall \onesigma\ directional
uncertainty is $\sim 35\degs.$ 
These results were obtained
using a maximum likelihood algorithm that minimizes
selection and Malmquist biases. Residual biases,
which have to do with the
geometry of the sample and the fact that
Cartesian velocity component errors add
in quadrature, were calibrated using Monte-Carlo simulations.
These showed 
that the raw maximum-likelihood flow amplitude
is biased high by \sm 25\%. The results quoted above
are corrected for this bias. The simulations
also enabled us to estimate that the probability
the survey would yield the derived flow vectors
by chance, if in reality the Hubble flow has converged
to the CMB frame at distances $\simlt 100\h1$ Mpc,
is 5.3\% for the ETR and 2.9\% for the full sample.

The LP10K flow is similar in amplitude and
scale to the bulk motion found
by Lauer \& Postman (1994) from their analysis
of 119 brightest cluster galaxies within 15,000 \kms.
However, the directions of the LP10K and LP94 flow vectors
differ by \sm 70\degs. When the LP10K likelihood analysis
is done with the flow required to lie along the LP94
direction, $l=343\degs,$ $b=52\degs,$ the best-fit
flow amplitude is only 150--200 \kms, much smaller
than the 700 \kms\ found by LP94, and the fit
quality is much worse
than for the velocity vectors quoted
in the previous paragraph. Thus, the LP10K
data set is inconsistent with the flow vector
measured by LP94.

The LP10K flow vector is similar in amplitude and
direction to the motion of
the Local Group with respect to the CMB, and lies
within \sm 40\degs\ of the flows measured in
the local ($cz\simlt 5000\ \kms$) universe in the
late 1980s and early 1990s by Lynden-Bell \etal\ (1988),
Willick (1990), Mathewson \etal\ (1992), and Courteau \etal\ (1993),
as well as thos obtained from a recent POTENT
analysis of the Mark III and SFI data sets by Dekel \etal\ (1998).
This suggests that the local universe, including the LG,
participates in the large-scale flow 
detected by the LP10K TF sample,
which in turn is driven by density 
inhomegeneities on scales $\simgt 150\h1$ Mpc.
The fact that the LP10K full sample yields the same
result as the ETR may hint that convergence to the CMB is not
seen even at distances $\simgt 200\h1$ Mpc.

The results presented in this paper do not, by themselves,
clinch the case for the reality of very large scale bulk flow,
because their significance level is only slightly above
$2\sigma.$ However, the excellent agreement of the LP10K TF
bulk flow with that recently measured by Hudson \etal\ (1998)
from the SMAC survey of elliptical galaxies adds considerable
weight to their plausibility. On the other hand, the LP10K and
SMAC results differ sharply from those reported in recent
months by Giovanelli \etal\ (1998a,b) and Dale \etal\ (1998),
who argue from the SFI/SCI TF data set that convergence
to the CMB frame is clearly detected beyond 60\h1\ Mpc. 
Resolution of this discrepancy is an important task for the near future.

The parameters
of the multiparameter TF relation introduced in
Paper I were solved for simultaneously with
the flow parameters, with the results largely unchanged.
In particular, a surface brightness dependence
of the TF relation was confirmed, with $v_{rot} \propto I_e^{0.13} L^{0.29}.$
The analysis in this paper treated the various sources of
TF scatter more carefully than did Paper I, with the result
that the instrinsic scatter of the TF relation is more
strongly constrained. It is found to be $\sigma_I = 0.28 \pm 0.07$ mag,
a value which should be accounted for by successful models
of galaxy formation.

\acknowledgements
The author acknowledges the support of NSF grant AST-9617188
and the Research Corporation, and thanks Paul Steinhardt,
Mike Hudson, and Keith Thompson for valuable discussions.


\begin{thebibliography}{}
\bibitem[]{} Abell, G.O., Corwin, H.G., \& Olowin, R.P.\ 1989, \apjs, 70, 1 (ACO)
\bibitem[]{} Bernstein, G.M., Guhathakurta, P., Raychaudhury, S., 
Giovanelli, R.,
Haynes, M.P., Herter, T., \& Vogt, N.P.\ 1994, \aj, 107, 1962
\bibitem[]{} Burstein, D., \& Heiles, C.\ 1978, \apj, 225, 40 (BH)
\bibitem[]{} Burstein, D., \& Heiles, C.\ 1982, \aj, 87, 1165
\bibitem[]{} Courteau, S., Faber, S.M., Dressler, A., \& Willick, J.A. 1993,
\apj, 412, L51
\bibitem[]{} Dale, D.A., Giovanelli, R., Haynes, M.P., Scodeggio, M.,
Hardy, E., \& Campusano, L.E.\ 1997, \aj, 114, 455
\bibitem[]{} Dale, D.A., Giovanelli, R., Haynes, M.P., Campusano, L.E.,
Hardy, E., \& Borgani, S.\ 1998, \apjl, in press (astro-ph/9810467)
\bibitem[]{} Dekel, A., Eldar, A., Kolatt, T., Yahil, A., Willick, J.A.,
Faber, S.M., Courteau, S., \& Burstein, D.\ 1998, preprint (astro-ph/9812197)
\bibitem[]{} Dressler, A., Faber, S.M., D., 
Burstein, D., Davies, R.L., Lynden-Bell,
D., Terlevich, R.J., and Wegner, G. 1987, \apjl, 313, L37
\bibitem[]{} Giovanelli, R., Haynes, M.P., Wegner, G., da Costa, L.N.,
Freudling, W., \& Salzer, J.J. 1996, \apjl, 464, L99
\bibitem[]{} Giovanelli, R., Haynes, M.P., Freudling, W., da Costa, L.N., Salzer, J.J., \&
Wegner, G.\ 1998a, \apjl, 505, L91
\bibitem[]{} Giovanelli, R., Haynes, M.P., Salzer, J.J., Wegner, G., da Costa, L.N., \&
Freudling, W.\ 1998b, \aj, in press (astro-ph/9808158) 
\bibitem[]{} Han, M.-S., \& Mould, J.R. 1992, \apj, 396, 453
\bibitem[]{} Holzapfel, W.L, Ade, P.A.R., Church, S.E., Mauskopf,
P.D., Rephaeli, Y., Wilbanks, T.M., \& Lange, A.E.\ 1997, \apj, 481, 35
\bibitem[]{} Hudson,~M.J., Smith, R.J., Lucey, J.R., Schlegel, D.J.,
\& Davies, R.L.~1998a, in {\em Evolution of Large-Scale Structure:
from Recombination to Garching,} Proceedings of the MPA/ESO Cosmology
Conference 
\bibitem[]{} Hudson, M.J., Smith, R.J., Lucey, J.R., Schlegel, D.J.,
\& Davies, R.L.\ 1998b, \apjl, submitted
\bibitem[]{} Jorgensen, I., Franx, M., \& Kjaergaard, P.\ 1996, \mnras, 280, 167
\bibitem[]{} Kogut, A.\ \etal\ 1993, \apj, 419, 1
\bibitem[]{} Lauer, T.R., \& Postman, M.\ 1994, \apj, 425, 418 (LP94)
\bibitem[]{} Lynden-Bell, D., Faber, S.M., Burstein, D., Davies, R.L, 
Dressler, A.,Terlevich, R., \& Wegner, G. 1988, \apj, 302, 536
\bibitem[]{} Mathewson, D. S., Ford, V. L, \& Buchhorn, M. 1992, \apjs,
81, 413
\bibitem[]{} Paczynski, B., \& Piran, T.\ 1990, \apj, 364, 341
\bibitem[]{} Perlmutter, S., \etal\ 1998, \apj, in press (astro-ph/9812133)
\bibitem[]{} Postman, M., \& Lauer, T.R.\ 1995, \apj, 440, 28
\bibitem[]{} Postman, M., Huchra, J.P., \& Geller, M.J.\ 1992, \apj, 383, 404
\bibitem[]{} Riess, A.G., Press, W.H., \& Kirshner, R.P.\ 1995, \apjl, 445, L91
\bibitem[]{} Riess, A.G., Davis, M., Baker, J., \& Kirshner,
R.P. 1997, \apj, 488, L1
\bibitem[]{} Riess, A.G., \etal\ 1998, \aj, 116, 1009
\bibitem[]{} Saglia, R.P., Colless, M., Burstein, D., Davies, R.L.,
McMahan, R.K., Watkins, R., \& Wegner, G.\ 1998, in 
{\em Evolution of Large-Scale Structure:
from Recombination to Garching,} Proceedings of the MPA/ESO Cosmology
Conference 
\bibitem[]{} Schlegel, D., Finkbeiner, D.P., \& Davis, M.\ 1998, \apj, 500, 525 (SFD)
\bibitem[]{} Steinmetz, M., \& Navarro, J.\ 1998, preprint (astro-ph/9808076)
\bibitem[]{} Strauss, M.A., Cen, R., Ostriker, J.P., Lauer, T.R.,
\& Postman, M.\ 1995, \apj, 444, 507
\bibitem[]{} Strauss, M. A., \& Willick, J. A. 1995, Phys.~Rep., 261, 271
\bibitem[]{} Struble, M.F., \& Rood, H.J.\ 1991, \apjs, 77, 363
\bibitem[]{} Wegner, G. \etal, 1996, \apjs, 106, 1
\bibitem[]{} Wegner, G. \etal, 1998, \mnras, in press
\bibitem[]{} Willick, J. A. 1990, \apj, 351, L5
\bibitem[]{} Willick, J. A., Courteau, S., Faber, S. M., Burstein, D., \&
Dekel, A. 1995, \apj, 446, 12
\bibitem[]{} Willick, J. A., Courteau, S., Faber, S. M., Burstein, D.,
Dekel, A., \& Strauss, M. A. 1997a, \apjs, 109, 333
\bibitem[]{} Willick, J.A., Strauss, M.A., Dekel, A., \&
Kolatt, T. 1997b, \apj, 486, 629
\bibitem[]{} Willick, J.A. 1999a, \apj, 516, 000, in press (Paper I; astro-ph/9809160)
\bibitem[]{} Willick, J.A. 1999b, in preparation (Paper III)
\end{thebibliography}
\end{document}